# Lunar ore geology and feasibility of ore mineral detection using a far-IR spectrometer

**Jakub Ciazela[1*], Jaroslaw Bakala[2], Miroslaw Kowalinski[2], Bartosz Pieterek[3], Marek Steslicki[2], Marta Ciazela[1], Grzegorz Paslawski[1], Natalia Zalewska[2], Lukasz Sterczewski[4], Zaneta Szaforz[2], Mateusz Jozefowicz[5], Dariusz Marciniak[1], Maciej Fitt[1], Adam Sniadkowski[6], Mirosław Rataj[2], Tomasz Mrozek[2]**

[1]Institute of Geological Sciences, Polish Academy of Sciences, Wrocław, Poland

[2]Space Research Centre, Polish Academy of Sciences, Warszawa, Poland

[3]Institute of Geology, Adam Mickiewicz University, Poznań, Poland

[4]Faculty of Electronics, Photonics and Microsystems, Wroclaw University of Science and Technology, Wrocław, Poland

[5]European Space Foundation, Kraków, Poland

[6]SKA Polska, Warszawa, Poland

**\*Correspondence:**
Jakub Ciazela
j.ciazela@twarda.pan.pl

**Abstract**

Lunar sulfides and oxides are a significant source of noble and base metals and will be vital for future human colonies' self-sustainability. Sulfide detection (pyrite and troilite) applies to many technological fields and use cases, for example, as a raw material source (available in situ on the Lunar surface) for new solar panel production methods. Ilmenite is the primary iron and titanium ore on the Moon and can provide helium-3 for nuclear fusion and oxygen for rocket fuel. The most important ore minerals have prominent absorption peaks in a narrow far-infrared (FIR) wavelength range of 20–40 μm, much stronger than the spectral features of other common minerals, including significant silicates, sulfates, and carbonates. Our simulations based on the linear mixing of pyrite with the silicates mentioned above indicated that areas containing at least 10–20% pyrite could be detected from the orbit in the FIR range. MIRORES, Multiplanetary far-IR ORE Spectrometer, proposed here, would operate with a resolution down to <5 m, enabling the detection of areas covered by 2–3 m$^2$ of pyrite (or ilmenite) on a surface of ~17 m$^2$ from an altitude of 50 km, creating possibilities for detecting large and local smaller orebodies along with their stockworks. The use of the Cassegrain optical system achieves this capability. MIRORES will measure radiation in eight narrow bands (0.3 μm in width) that can include up to five bands centered on the ore mineral absorption bands, for example, 24.3, 24.9, 27.6, 34.2, and 38.8 μm for pyrite, marcasite, chalcopyrite, ilmenite, and troilite, respectively. The instrument size is 32 x 32 x 42 cm, and the mass is <10 kg, which fits the standard microsatellite requirements.

# 1   Introduction

The Earth notices a growing demand for specific elements necessary to transition into a zero-carbon economy and energy production (International Energy Agency, 2022). The supply and ore grade of these metals on Earth is gradually declining. Many elements in ore minerals are critical metals for which demand already exceeds (or is expected to exceed) supply. Critical metals are crucial for the development of advanced technologies and clean energy. The most outstanding example is undoubtedly the elements necessary for the transition to low-carbon technologies and the electrification of transport. In addition, critical metals are found in superalloys and other high-tech medical, aerospace, and telecommunications devices. Space mining also needs critical metals on Earth to develop some spaceship and space habitat infrastructure to be sent into space. More importantly, local resources of essential metals will be required on the Moon to avoid costly transportation from Earth. Due to the high cost, importing iron and copper from the Moon will be impractical. However, their *in-situ* presence for future human activities will save the Earth's resources for climate action instead of sending them to space in large amounts for exploration and other purposes. Importing some REEs, cadmium, and cobalt to Earth might still be feasible if it is necessary for specialized electronics and battery production. The most likely import from the Moon is helium-3 for emerging nuclear fusion.

Among the four strategic resources envisaged at the stage of exploration of the Solar System by humans, apart from water and propellant fuels, the European Space Agency (ESA) listed base metals (e.g., Fe, Ti, Cu) and precious metals (e.g., platinum group elements (PGE), Au, Ag) (ESA, 2019). Some chalcophile and siderophile metals are expected to be critical to the future colonization of the Moon. Metals, especially those chalcophile, are crucial in producing electrical wires. Traditionally, copper was used for this, but modern devices that require high-speed signal conduction, such as desktop computers, laptops, and mobile phones, use gold or silver. Critical metals will have to be mined in situ in the long term, as systematically transporting large loads between the Earth and the Moon would be highly costly financially and environmentally. Chalcophile and siderophile metals are found mainly in sulfides and oxides and sometimes in platinum group minerals (PGM) and native metals, which are the primary sources of most precious metals (Au, Ag, Pt, and Pd) and base metals (Fe, Ti, V, Cr, Cu, Pb, Zn, Sn, Co, Ni, and others) on Earth (U.S. Geological Survey, 2016). Besides oxides, sulfides are the most critical group of ore minerals (Vaughan and Corkhill, 2017) and are essential for the recovery of Cu, Ag, and Au (Ciazela et al., 2017). Chalcophile metals, such as Ag, Cu, Zn, and Pb, are found primarily in sulfides. Siderophile metals can occur in various minerals, mainly oxides and sulfides. These include Fe, Ti, Cr, Mn, V, Pt, and Pd. Oxides and sulfides usually coexist on the Moon (Hu et al., 2021; Saal and Hauri, 2021).

Despite the growing interest in space mining of asteroids and planetary bodies, directly detecting sulfides and oxides from orbits was an unexplored field as they were undetectable from orbits. However, far-infrared (FIR) spectroscopy may offer rich possibilities for detecting them from orbit. On Earth, these ore minerals are the primary source of industrial and precious metals and play an essential role in mining activities. However, they were searched on Earth using traditional geological mapping, drilling, and geophysical or remote sensing methods, including ground penetrating radar (GPR), magnetometry, gravimetry, and near-infrared spectroscopy. In space, the most advanced infrared spectrometer, the Compact Reconnaissance Imaging Spectrometer for Mars (CRISM), launched in 2005 on the Mars Reconnaissance Orbiter (MRO) spacecraft, is not able to detect sulfides, but other selected minerals (mostly sulfates, such as jarosite and alunite) which can be used as a proxy in the search for sulfides. There is not even an instrument of this class on the Moon yet.

Therefore, we propose a lunar version of the MIRORES (Multiplanetary far-IR Ore Spectrometer) instrument developed from its Martian version (Ciazela et al., 2022). MIRORES should be able to detect metal resources in sulfides and oxides also on Mars (Ciazela et al., 2022), where the low concentration of water in the atmosphere makes excellent conditions for FIR measurements, but the Moon is the most favorable environment for MIRORES, due to its extremely thin atmosphere, being technically just an exosphere. Therefore, MIRORES, which might be the world's first high-resolution far-infrared spectrometer for planetary research, will map various Moon resources during the Polish Lunar Mission. The mission based on the MIRORES instrument envisages the selection of areas on the Moon with the most significant resource potential regarding iron, titanium, copper, noble metals, sulfur, and helium-3. These will be areas of strategic importance for future lunar bases. Preparations for the mission will begin with Phase A in the fall of 2023, and the mission should last from 2028 to 2032.

FIR remote sensing mainly applies to terrestrial planets with a thin or no atmosphere. On Earth, FIR radiation received by satellites is primarily attenuated by water vapor in the Earth's atmosphere. Water, which is a strong absorbent in the Earth's atmosphere, containing ~40 kg·m$^{-2}$ of water in the atmospheric column, is no longer a problem on Mars (only 0.01–0.1 kg·m$^{-2}$ in the atmospheric column), and on the Moon, water is absent (Read et al., 2015). In addition, the Moon is an easier-to-reach exploration target than Mars and may enable *in situ* resource utilization (ISRU) policy (Valentini et al., 2022).

## 2      Geology of the Moon

The Moon is a differentiated celestial body with a core, mantle, and crust (Wieczorek et al., 2006). Because the Moon has only an extremely thin atmosphere (Benna et al., 2015), which eliminates weather-induced erosion, the geomorphology of the Moon's surface has mainly been shaped by magmatism and space weathering. Space weathering includes solar wind and radiation, cosmic rays, micrometeorites, and large meteorite impacts (Jaumann et al., 2012). The most striking feature of the Moon is the contrast between its dark and light zones. The dark zones correspond to lunar maria, mostly 3.0–4.0 Ga old (Meyer, 2003; Kring and Durda, 2012), often filling older impact craters. The bright zones represent an old lunar crust from the differentiation of the primary magma ocean, being mostly >3.9 Ga old (Meyer, 2003; Kring and Durda, 2012) and of anorthositic composition, referred to as the highlands. Typical lunar mare basalts have comparatively high levels of FeO$_T$ (>16 wt%), low to moderate levels of MgO, high levels of TiO$_2$, and low levels of Al$_2$O$_3$ (Warren and Taylor, 2013). Lunar mare basalts can be classified according to their TiO$_2$ content into high TiO$_2$ basalts (>6 wt% TiO$_2$), low to medium TiO$_2$ basalts (1.5–6.0 wt% TiO$_2$), and very low TiO$_2$ basalts (<1.5 wt% TiO$_2$) (Shearer and Papike, 1999). All three types are composed mainly of common silicates, i.e., olivine (fayalite and forsterite), pyroxene (mainly pigeonite), and plagioclase (anorthite to bytownite), while the other minerals, including oxides and sulfides, are accessory minerals (Papike et al., 1991; Shearer and Papike, 1999). Basalts with high TiO$_2$ content typically consist of olivine (~5 vol.%), pyroxene (~50 vol.%), plagioclase (~30 vol.%), and opaque minerals, i.e., oxides and sulfides (~15 vol.%) (Papike et al., 1976). Lunar mare basalts contrast strongly with the rest of the Moon's crust. The typical composition of the highlands, dominated by anorthosites (>90 vol.% plagioclase), averages 5.0 wt% FeO$_T$, 0.4 wt% TiO$_2$, and 27 wt% Al$_2$O$_3$ (Warren and Taylor, 2013). Based on bulk-analysis data, Wood (1975) proposed using TiO$_2$ content and Ca/Al ratio to discriminate mare basalts and highland material.

The most lunar surface is covered by a few-meter-thick regolith, a product of space weathering and erosion of the Moon's rocky surface (Fa and Jin, 2010). Therefore, lunar regolith consists of fragments of igneous intrusive and extrusive rocks, but also various types of breccias produced by meteoroid impacts, impact glass, and agglutinates, which are glass-welded grain aggregates (Lucey et al., 2006). Breccias are mostly polymict—they contain clasts from many older rocks bound with the matrix, whereas some are dymict or monomict (Lucey et al., 2006). Their composition differs depending on the location, with more felsic clasts in the lunar highlands and more mafic clasts in the lunar maria. Space weathering includes collision with small particles and tiny meteorites that affect the lunar surface by an increasing maturation of the regolith material by (1) lithification, (2) mechanical comminution, (3) melting and sublimation, (4) formation of agglutinates, (5) implantation of ions, (6) sputtering, and (7) incorporation of meteoritic material (Jaumann et al., 2012). Micrometeorites, in particular, cause abrasion and fragmentation of regolith, in contrast to larger impactors, which may involve heating and local or regional melting of rocks (Jaumann et al., 2012).

The bedrock's mineral and chemical composition primarily determines the regolith's composition. Therefore, it varies depending on the region in which it occurs. Regolith covers the lunar surface unevenly. The regolith cover is thinner on the Moon's near side than on the far side. It varies from about 2 to 14 m. Most often, however, the thickness of the regolith is from 2 to 8 m, only locally >8 m (Fa and Jin, 2010) (Figure 1). The regolith cover is thicker on its far side, mainly between 6–10 m, but occasionally even >14 m. It should also be noted that the circumpolar regions are covered with the thinnest regolith layer, with a 2–4 m thickness, reaching a maximum of ~6 m (Fa and Jin, 2010) (Figure 1).

The optimal inclination of the MIRORES orbit (see Section 7), which would be close to 90°, means that the circumpolar regions should be studied in detail. Most of the North Arctic Circle (>65°N) is covered by lunar highlands, and only ~10%, mainly in the South East South (SES), is covered by the Frigoris Mare Basin Margin Formation. The anorthosites predominating lunar highlands are mostly Aitkenian (4.30–3.92 Ga) in the west and Nectarian (3.92–3.85 Ga) east (Ji et al., 2022). Only crater deposits show younger ages, which are, however, rare. Those of the Imbrian age (3.85–3.16 Ga) are sparsely but somewhat regularly distributed throughout the area (Ji et al., 2022). Only a few craters are Eratosthenian (3.16–0.80 Ga) or Copernican (<0.80 Ga), and they are located only in the south (Ji et al., 2022).

In the south, near the South Pole, the formations of the South Polar Circle are similar to those of the North Polar Circle, covered mainly with Aitkenian and Nectarian anorthosites with a tiny proportion of Imbrian, Eratosthenian, and Copernican crater deposits (Ji et al., 2022). However, in the north of the South Polar Circle, some areas are covered on one side by the magnesium anorthosite complex (Ji et al., 2022) and on the other by sediments of the Aitken Basin, which is a vast area associated with a high gravity anomaly and a shallow crust-mantle boundary (James et al., 2019).

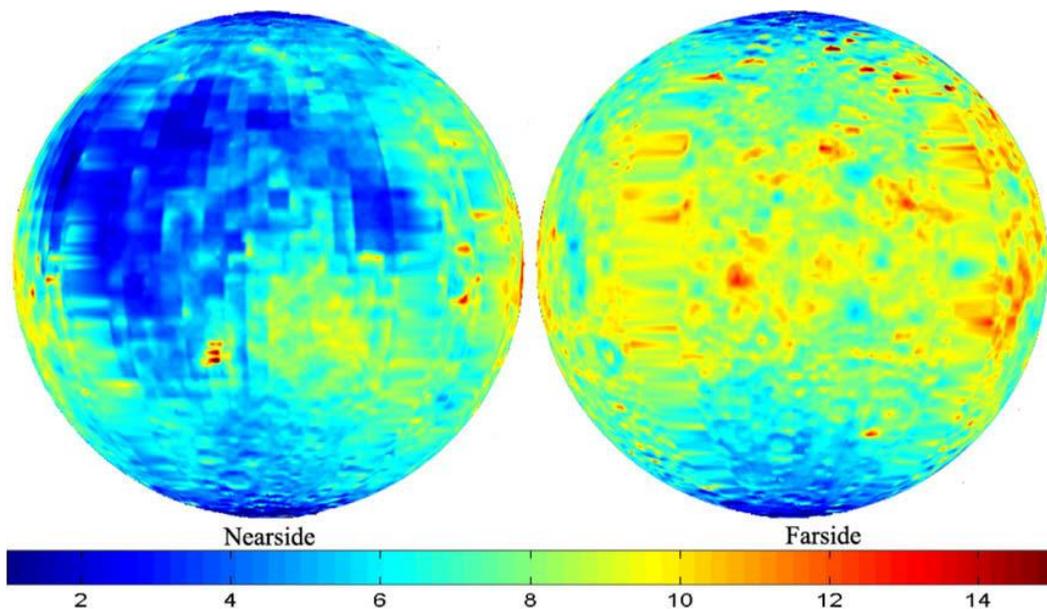

**FIGURE 1.** Global thickness (in meters) distribution of the regolith cover on the Moon. Basalt mare and circumpolar areas are covered with the thinnest layer of regolith. Redrawn from Fa and Jin (2010).

## 2.1 Rocks of the Apollo program

Between 1969 and 1972, six crewed missions to the Moon were carried out as part of the Apollo program (Hiesinger and Head, 2006), with later programs including soviet Luna (Hiesinger and Head, 2006) and Chinese Change (Li et al., 2022a) being uncrewed. Although >50 years have passed since the last human presence on the Moon, the 2020s should bring a breakthrough thanks to the Artemis program. Artemis 1 mission, a 25-day flight around the Moon, was completed on December 11, 2022 (Gee et al., 2023). Artemis 2 will perform a similar operation 2024 with a crew (David et al., 2023). During the Artemis 3 mission, the crew will descend to the lunar surface, and the multi-year construction of the lunar base will begin (Boazman et al., 2022). From now on, we will need lunar resources. Of the 384 kg of rock material brought to Earth from space missions to the Moon, 382 kg were fetched by the Apollo program, and only 2 kg by the Soviet Luna program (1970–1976) and the Chinese mission Change-5 (2020) (Li et al., 2022b). Each successive Apollo mission brought more and more material to Earth: from 22 kg for the first Apollo 11 mission to 110 kg for the last Apollo 17 mission (Hiesinger and Head, 2006). Apollo 16 explored the lunar highlands, Apollo 11 and 12 the lunar mare, while Apollo 14, 15, and 17 the lunar mare and highland boundaries (Hiesinger and Head, 2006). None of the programs covered the polar regions, although some rocks collected in the Apollo program resemble those typical of polar regions (Doyle, 2017).

Apollo 11 collected mainly basalts and breccias. Small but significant fragments of anorthositic crust from the lunar highlands were found in some Apollo 11 breccias and interpreted as evidence of an early magma ocean on the Moon (Fuller, 2014). In contrast, Apollo 12's rocks were almost entirely basalt, with only two breccias in the brought samples. Basalts are composed mainly of plagioclase and pyroxene, with additions of olivine and ilmenite. As would be expected in a region formed by the debris of an impact basin, most of the 42 kg of rock and soil collected on Apollo 14 were breccias (Hiesinger and Head, 2006). Many of these breccias contain material enriched in KREEP, a group of incompatible elements that includes potassium (K), rare earth elements (REE), and phosphorus (P). KREEP is thought to have formed in the first phase of the Moon's history, during the solidification of the Moon's molten phase known as the magma ocean (Jolliff et al., 2000).

Apollo 15 also collected breccias, but in addition, basalts and lunar highland rocks (anorthosites). Breccias dominated the Apollo 16 sample collection, which collected several important crustal rock types of the lunar highlands areas. The Apollo 17 crew collected all types of rock previously collected by Apollo 15 and several rarer types of lunar rocks representing the middle and lower lunar crust, including norite, troctolite, and dunite (Hiesinger and Head, 2006).

## 3 Potential resources of the Moon

Ilmenite ($FeTiO_3$) is the best source of Fe and Ti on the Moon. Iron and titanium, as well as sulfur as the main component of sulfur concrete (Grugel and Toutanji, 2008), will be used to build edifices. Metals will be the resources for building structures on the Moon through additive manufacturing based on powders or by sintering them. Copper, Ag, and Au in sulfides will be conductors in modern electronic equipment. Pyrite or other sulfides converted to pyrite can be used to make solar panels (Kristmann et al., 2022).

On the Moon, besides metals, helium-3 and other gases (which can be used as propulsion fuels) can also be prospected. Volatile entrapment on the Moon has been occurring primarily due to the specific crystallographic structure of ilmenite, which captures large amounts of $^3He$ from the solar wind (Kim et al., 2019). This fact is known from the analysis of Apollo samples, which revealed a high correlation between the content of $^3He$ and $TiO_2$ (Johnson et al., 1999 and Figure 2). Ilmenite is, therefore, a good source of this He isotope. The demand for $^3He$ is caused by the possibility of using it in fusion reactors, which will be the cheapest and safest energy source. ESA is already taking steps to prepare special thermoses for transporting $^3$He from the Moon to Earth (https://ideas.esa.int/servlet/hype/IMT?documentTableId=45087631481217827&userAction=Browse&templateName=&documentId=785a1b7b1d1eae6cb66829f14d4e4f6f). Oxygen, which can also be obtained from ilmenite as a by-product of Fe and Ti extraction (Lomax et al., 2020), together with $H_2$ will serve as rocket fuel for Earth-return missions and in life support systems. ESA invests in contracts to extract $O_2$ from lunar soil using two chemical processes: FFC Cambridge (abbreviation from the creators: Fray, Farthing, and Chen) and reduction of ilmenite with $H_2$ (https://www.spaceapplications.com/news/esa-awards-space-resources-contract-to-space-applications-service). A process that allows both oxygen and metals to be produced from ilmenite has been reported by Schwandt et al. (2012). Further works focus on separating various gases obtained from ilmenite or regolith (Schlüter et al., 2021), helium from other gases (Janocha, 2020), and $^3$He from $^4$He (Kempiński et al., 2019). Therefore, ilmenite is currently being studied as a resource (Johnson et al., 1999; Fa and Jin, 2007; Kim et al., 2019).

Ilmenite can be expected, especially in lunar mare (~20% of the Moon's surface), accounting for an average of ~4 wt% of the minerals. Ilmenite is suggested by the elevated concentrations of Fe and Ti estimated by the Clementine mission (Lucey et al., 1998) (Figure 3). The MIRORES device aims to map the areas with the highest concentrations of ilmenite (>15 wt%) and other oxides and sulfides in more detail. Landing in the most ilmenite-rich location will be critical for developing future human and industrial bases. Therefore, the MIRORES data may be of great scientific and economic value.

While the amounts of ilmenite on the Moon are vast, finding sulfide ore deposits is difficult. Despite the lack of direct evidence, we assume sulfide ore deposits occur on the Moon due to sulfides in lunar meteorites and lunar rock samples brought back to Earth. For example, Koeberl et al. (1996) found numerous Fe sulfides in the lunar meteorite QUE 93069. Recently, ilmenite with troilite was found in a piece of CE5 basalt collected by Change-5 (Hu et al., 2021). The co-occurrence of sulfides with Fe-Ti oxides is confirmed by the excellent correlation between the S and Ti contents in lunar samples (Saal and Hauri, 2021).

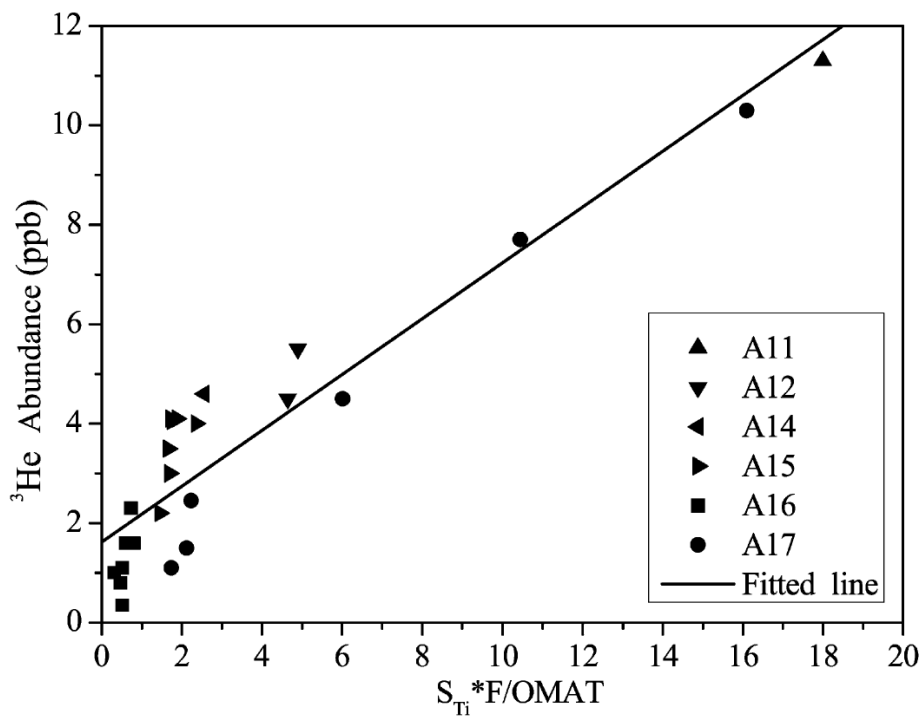

**FIGURE 2.** Helium-3 content vs. the product of the $TiO_2$ content ($S_{Ti}$) and the modeled solar wind flux (F) divided by optical maturity (OMAT) in 21 regolith samples from the Apollo missions from 11 to 17. The correlation coefficient is 0.94. Redrawn from Fa & Jin et al. (2007).

However, detecting ore minerals on the Moon has been challenging based on current orbital instruments. In the first era of lunar exploration related to the Cold War, from the Soviet Luna 3 mission in 1959, through the American Ranger program (1961–65), to the Apollo program (1961–72), the focus was solely on the photographic documentation of the Moon (Pędzich and Latuszek, 2014). It was only in a new era, from the 1990s, that advanced spectrometers began to be used in NASA missions. The Clementine probe launched in 1994 was equipped with a laser altimeter and four small-format CCD cameras for observing and mapping the Moon. The cameras have gathered images covering the entire surface of the Moon in five spectral ranges (Pędzich and Latuszek, 2014), which made it possible to carry out a general recognition of its chemical composition (Figure 3). Lunar Prospector orbited the Moon above its polar areas in 1998 and 1999. The probe was equipped with gamma, neutron, and alpha radiation spectrometers to determine what elements the lunar surface is composed of (Pędzich and Latuszek, 2014).

Apart from these two missions, the most critical geochemical data was provided by the Indian probe Chandrayaan-1 launched in 2008, equipped with the Moon Mineralogy Mapper ($M^3$) spectrometer operating in the wavelength range from 0.4 to 3.0 μm and a resolution of 70 m/pixel and a similar spectrometer SIR-2 launched in 2008 (Green et al., 2011) operating with a low resolution of ~220 m/pixel (Bugiolacchi et al., 2011). Even spectra of $M^3$ allow for an accurate but only large-scale interpretation of the mineralogy of the studied areas (Figure 4) (Green et al., 2011). Moreover, while lunar rovers have found significant sulfides and more were found in lunar meteorites with their spectra available from the U.S. Geological Survey (USGS) spectral library (https://speclab.cr.usgs.gov/spectral-lib.html, access December 15, 2022), the poor near-infrared (NIR) spectral characteristics of sulfides and interference with pyroxenes make it challenging to detect sulfides in the NIR range available in $M^3$. For example, pyrite shows a flat increase in spectral values >1 μm, which resembles clinopyroxene (Figure 5). Their similarity is critical because clinopyroxene (primarily pigeonite) is one of the few most abundant minerals on the lunar surface, especially in lunar mare basalts (Papike et al., 1976, 1991; Shuai et al., 2013).

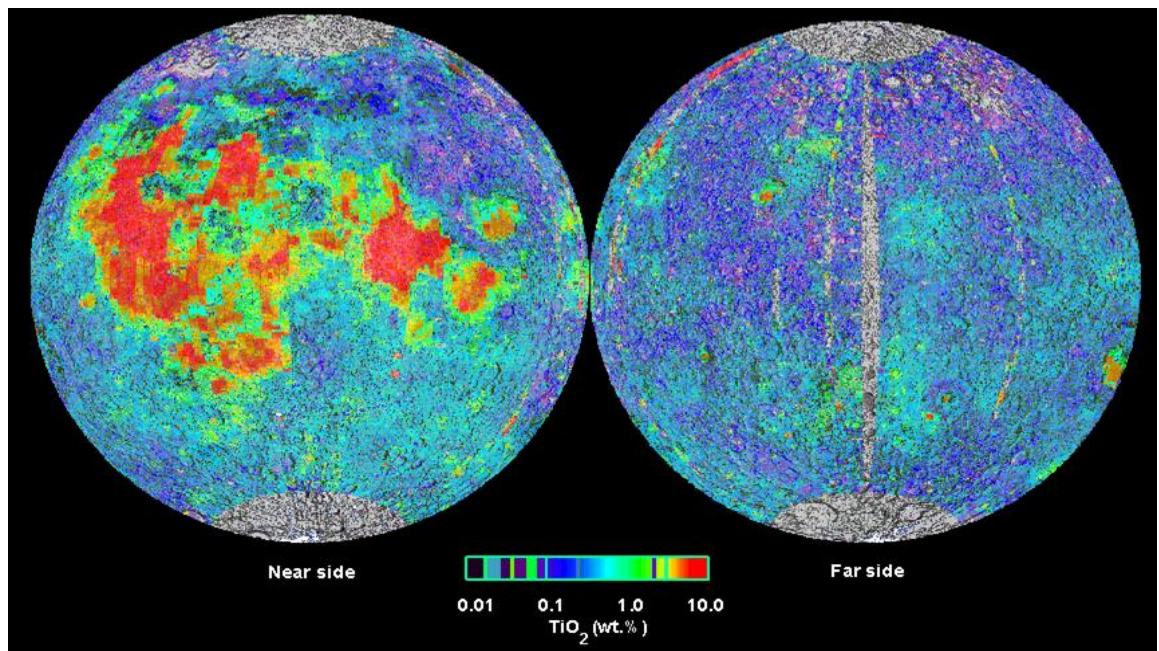

**FIGURE 3.** Image obtained from global Clementine data (at wavelengths of 415 and 750 nm) shows $TiO_2$ concentration in the regolith on the lunar surface. Noteworthy is the low level of $TiO_2$ in the anorthositic highlands and its high content in basalt mare basins (visible in the left image), especially Oceanus Procellarum (west) and Mare Tranquillitatis (central east, site of the first man landing on the Moon), and to a lesser extent Mare Imbrium (central part). Redrawn from the Lunar and Planetary Institute.

On the other hand, common ore minerals, oxides, and sulfides on the Earth and the Moon (for example, ilmenite, troilite, and pyrite) have distinct absorption bands in the FIR range, between 20–40 μm (Figure 6). The same applies to other common sulfides, such as pyrrhotite (Hony et al., 2002) or chalcopyrite (Brusentsova et al., 2012). Notably, the spectral features of the ore minerals mentioned above in the FIR range do not significantly interfere with the rock-forming minerals (Figure 6), in contrast to the currently available NIR range (Figure 5). Trigonal quartz, which may also occur locally on the Moon (Fagan et al., 2014), shows one of its smaller peaks also in the FIR, at 401 cm$^{-1}$ (24.9 μm) (Farmer, 1974) close to the pyrite band, which, however, is outside the measured pyrite band at 409–414 cm$^{-1}$ (24.15–24.45 μm).

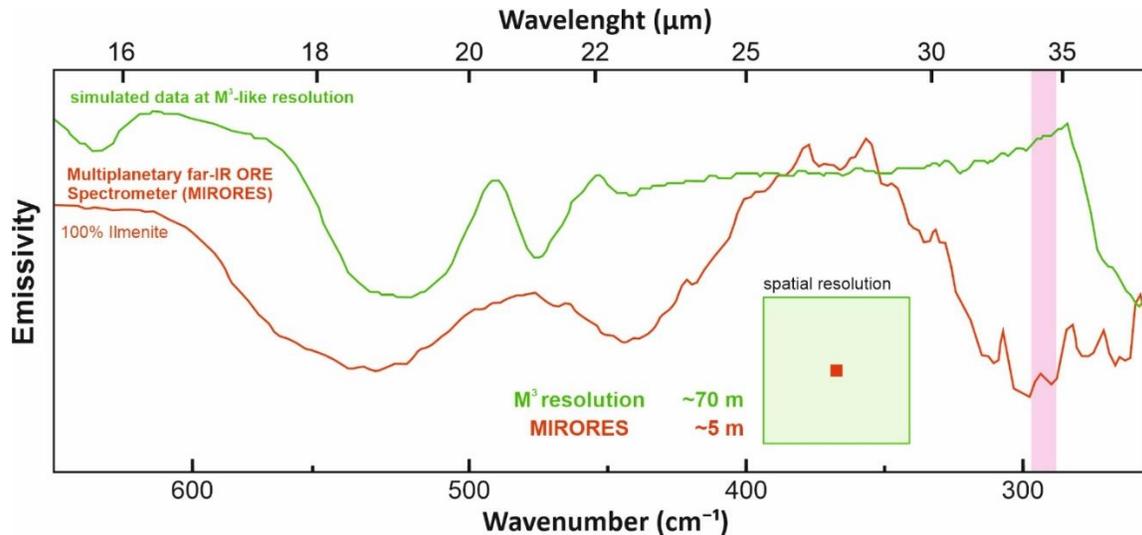

**FIGURE 4.** Our simulation of the observed emissivity spectra from a 5 x 5 m field covered with ilmenite and surrounded by an infinite area of lunar mare basalts (see Section 5 for the composition) as would be recorded by a far-infrared (FIR) instrument operating at a Moon Mineralogy Mapper (M$^3$)-like resolution and the MIRORES spectrometer. Simulations show that small and medium-sized orebodies would be impossible to detect with infrared spectrometers with the resolution of those previously orbiting the Moon, even if they could have operated in the FIR. The ordinate axis does not have a numerical scale because the graphs are shifted for clarity.

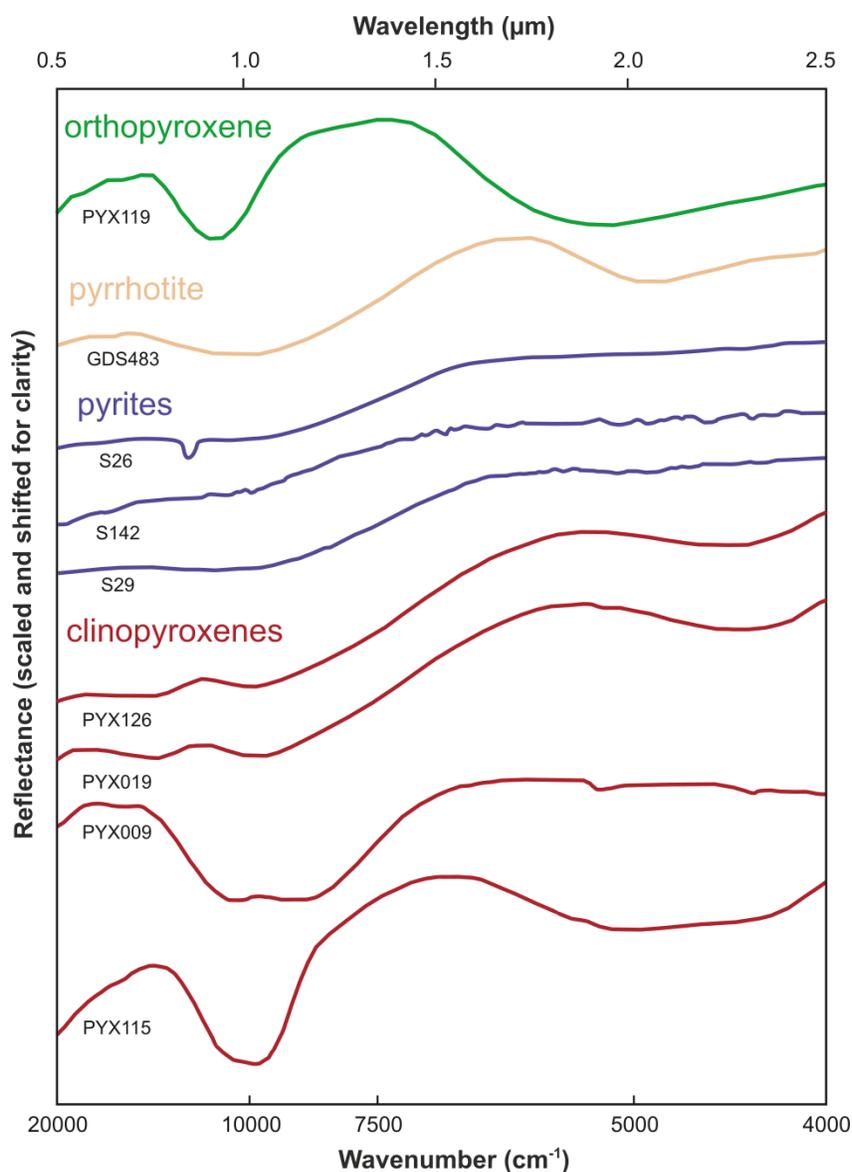

**FIGURE 5.** Near-infrared (NIR) spectra of pyrrhotite (USGS spectral library) compared to orthopyroxene (RELAB spectral database), as well as pyrites S26, S142, and S29 (USGS) compared to those of clinopyroxenes PYX126, PYX019, PYX009, and PYX115 (RELAB). Noteworthy is the lack of distinct spectral features of sulfides and their similarity to the NIR spectra of some pyroxenes. The low content of sulfides compared to pyroxenes makes sulfides challenging to observe on the Moon in the NIR range. The ordinate axis does not have a numerical scale because the graphs are shifted for clarity. Adapted from Ciazela et al. (2022) and Horgan et al. (2014).

## 4 Spectral ranges and interferences

The instrument will measure the FIR spectra of the lunar surface using detectors collecting signals from five narrow bands between 20–40 µm, where pyrite (24.3 µm = 412 cm$^{-1}$), marcasite (24.9 µm = 402 cm$^{-1}$), chalcopyrite (27.6 µm = 362 cm$^{-1}$), ilmenite (34.2 µm = 292 cm$^{-1}$), and troilite (38.8 µm = 258 cm$^{-1}$) have their prominent absorption bands. Depending on the mission needs, 2−3 additional detectors can be inserted into this system to measure up to 7–8 minerals simultaneously and include, for example, REE-rich apatite. The radiation received by the detectors depends on many additional factors that need to be normalized, including the Sun's geometry and the slope's inclination. Therefore, additional reference detectors collect the signal from the 20.7, 21.5, and 31.5 µm lines (483, 465, and 317 cm$^{-1}$), where sulfides and ilmenite do not show spectral features. Such an approach will allow determining the influence of the Sun's geometry and topography on the obtained results - a similar approach is applied in the Landsat 8 (Safari et al., 2018; Banerjee et al., 2019; Sekandari et al., 2020), ASTER (Safari et al., 2018; Zoheir et al., 2019; Sekandari et al., 2020), Sentinel-2 (Zoheir et al., 2019; Sekandari et al., 2020; Soydan et al., 2021), or PRISMA (Loizzo et al., 2018; Bedini and Chen, 2022; Chirico et al., 2023) spectral bands on Earth. Apart from minor interferences with $H_2O$ and $CO_2$ molecules in the FIR band, which in any case is nearly irrelevant on the Moon, the main FIR advantage is the robust absorption features of oxides and sulfides (Figure 6A) compared to rock-forming silicates (Figure 6B) and other common minerals (Figures 6C and 7).

The lunar atmosphere, technically an exosphere, contains only trace amounts of certain volatile elements near the lunar surface. These are primarily Ar with a surface area of 20,000−100,000 atoms per cm$^3$, He with 5,000−30,000 atoms, Ne with up to 20,000 atoms, Na with up to 70 atoms, K with 17, and H with <17 (Benna et al., 2015). Carbon dioxide exhibits a strong vibrational and rotational absorption band centered ~15 µm, but pure $CO_2$ absorbs relatively little beyond this band and is extremely rare in the lunar atmosphere (Klumov and Berezhnoi, 2002; Berezhnoy et al., 2003). Similarly, $H_2O$, a strong absorbent on Earth, with ~40 kg m$^{-2}$ in the Earth's atmosphere column, is only ephemeral in Moon's atmosphere. As with $CO_2$, it is mainly associated with comet impacts in the polar regions (Klumov and Berezhnoi, 2002; Berezhnoy et al., 2003). While $H_2O$ is not a problem in most of our target range between 20−40 µm, it can become an effective absorber at >35 µm and selected lines <35 µm. In general, however, the absorbance of the lunar exosphere in the FIR range is several orders of magnitude lower than the absorbance of the Earth's atmosphere. The Moon's exosphere is also relatively dust-free. Only near the surface, the Moon has a thin layer of moving dust particles that constantly bounce and fall to the surface, which is caused by electrostatic levitation (Stubbs et al., 2006). The effect seems to be stronger on the dark side and at the terminator separating the night and day sides of the Moon (Stubbs et al., 2006). On the bright, day side of the Moon, the dust effect is negligible for IR observations.

The main advantage of the FIR measurements on the Moon is the strong spectral features of sulfides and oxides in this range, which are much stronger than the spectral features of any of the common lunar silicates documented by the Apollo program or lunar meteorites, i.e., olivines, pyroxenes, and plagioclase (Figure 6) (Papike et al., 1976). The prominent spectral features of ore minerals in the FIR range are a significant advantage over the NIR range, where sulfides are almost indistinguishable from pyroxenes (Figure 5). Sulfates, which may also occur on the lunar surface (Mokhov et al., 2008), especially those non-hydrated, do not interfere with sulfides in FIR (Figures 6C and 7B). Akaganeite, on the other hand, shows an absorption band at 414 cm$^{-1}$ (24.2 µm) (Chukanov and Chervonnyi, 2016), similar to that of pyrite (Figure 7A). However, it is a rare mineral, occurring most often as an alteration of pyrrhotite (Carter et al., 2015). Its only detection so far in lunar rocks, in Apollo 16 samples, has been interpreted as a direct result of the oxidation of these samples caused by water vapor contamination that occurred after the probe returned to Earth. Therefore, the only akaganeite discovered in lunar rocks is probably of entirely terrestrial origin (Taylor and Burton, 1976). Interference with pyrite may also occur among arsenides; for example, mimetite (418 cm$^{-1}$ = 23.9 µm) (Chukanov and Chervonnyi, 2016). However, mimetite and other arsenides occur mainly in metal ore oxidation zones (Bajda, 2010). Therefore, the potential presence of mimetite would only increase the chances of finding sulfide mineralization.

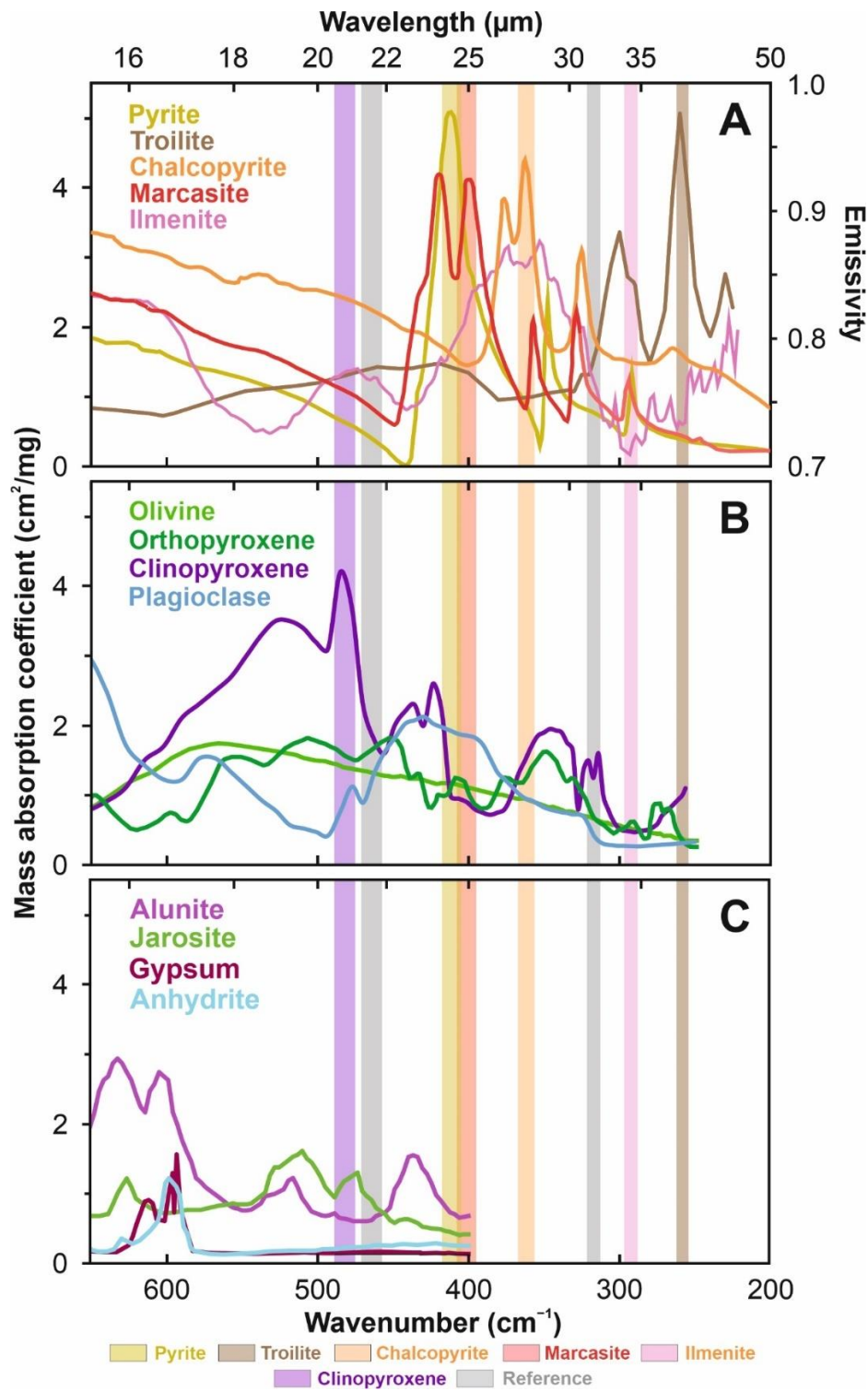

**FIGURE 6.** Mass absorption coefficients of possible lunar **(A)** ore minerals (Hony et al., 2002; Brusentsova et al., 2012), **(B)** silicates (Koike et al., 2000; Gielen et al., 2008; Chihara and Koike, 2017), and **(C)** sulfates (Bishop and Murad, 2005; Bishop et al., 2014; Bhattacharya et al., 2016) measured at room temperature. Literature data for sulfates and silicates are unavailable for <400 cm$^{-1}$ (>25 μm) and <250 cm$^{-1}$ (>40 μm), respectively. Sulfate mass absorption coefficients were calculated from data on transmittance, sample weight, and active cross-sectional area according to the method presented by Brusentsova et al. (2012). The mass absorption coefficient for troilite is normalized (troilite maximum = pyrite maximum) because Hony et al. (2002) only report normalized values without specifying normalization factors. The values for ilmenite represent the emissivity (see the right Y-axis) since to the best of our knowledge the mass absorption coefficients and their derivatives are not yet available in the literature. The data source on the emissivity of ilmenite is the Arizona State University spectral library (https://speclib.asu.edu/, ilmenite no. 463). Note that emissivities are not interchangeable with mass absorption coefficients, but low emissivity values mean high values of mass absorption coefficients. Therefore, we focus on the low-value interval of emissivity at the 34.2 µm (pale pink band), where the silicates **(B)** show low values of mass absorption coefficients (implying high emissivity in contrast to ilmenite). Although grain size distribution may affect mid-infrared features of minerals, the differences observed between large and small particle sizes in the far-infrared range (<550 cm$^{-1}$) seem to be negligible (Hamilton et al., 2020). The rectangles indicate the spectral ranges of the eight detectors: brown (254–262 cm$^{-1}$ = 38.22–39.38 µm: troilite), pale pink (288–297 cm$^{-1}$ = 33.69–34.71 µm: ilmenite), orange (357–368 cm$^{-1}$ = 27.19−28.01 µm: chalcopyrite), red (396–408 cm$^{-1}$ = 24.53−25.27 µm: marcasite), yellow (405–418 cm$^{-1}$ = 23.94–24.66 µm: pyrite), gray (313–322 cm$^{-1}$ = 31.03–31.97 µm and 458–472 cm$^{-1}$ = 21.18–21.82 µm: reference bands) and purple (476–490 cm$^{-1}$ = 20.39–21.01 µm: clinopyroxene).

Another ore mineral, hematite, probably more common than sulfides on the Moon (Li et al., 2020), has its main absorption line at 467 cm$^{-1}$ (21.4 µm), which is far from the sulfide lines (Figure 7A). Potassium-rich alkali feldspars, common in terrestrial granites, such as microcline with its absorption line at ~413 cm$^{-1}$ (24.2 µm), would make sulfide detection difficult (Figure 7C). Microcline, however, forms in relatively low temperatures in the presence of fluid (Perkins et al., 2023), so it is not expected on the Moon. Other alkali feldspars, orthoclase (~439 cm$^{-1}$) or sanidine (~443 cm$^{-1}$) (Salisbury et al., 1987), are less problematic for ore minerals. Besides, even if not weathered to clay minerals, they are associated with evolved silica-rich felsic magmas exotic on the Moon. In any case, they are restricted to areas other than basalt mare (Clegg-Watkins et al., 2017). Among the common secondary silicates, prehnite may be the only mineral interfering with the sulfide signal (Figure 7B). Prehnite, however, has not yet been found on the Moon. Prehnite shows weak absorption lines at 420 cm$^{-1}$ (23.8 µm) near the pyrite line, and thus more considerable amounts of prehnite can mimic the signal of smaller amounts of pyrite. However, at higher concentrations of sulfides (>10%), for which MIRORES is intended, the potential interference of prehnite with pyrite is negligible. The same may be true of carbonates. However, carbonates are extremely rare on the Moon. In addition, magnesium carbonates, magnesite, and dolomite exhibit broad but relatively shallow absorption characteristics superimposed on sulfides. These features are so broad that they extend to the reference detectors and are easily distinguishable from sulfides in the received signal (Figure 8C).

Based on those analyses, we designed a relatively inexpensive and straightforward FIR spectrometer based on pyroelectric detectors measuring eight narrow (0.6−1.2 µm) diffraction grating-split spectral bands, including five prominent bands centered at 24.3 µm (412 cm$^{-1}$) to detect pyrite, 24.9 µm (402 cm$^{-1}$) for marcasite, 27.6 µm (362 cm$^{-1}$) for chalcopyrite, 34.2 µm (292 cm$^{-1}$) for ilmenite, 38.8 µm (258 cm$^{-1}$) for troilite, and two reference bands centered at 21.5 µm (465 cm$^{-1}$) and 31.5 µm (317 cm$^{-1}$). The last band will monitor pyrite-clinopyroxene interference (cf. Ciazela et al., 2022) at 20.7 µm (483 cm$^{-1}$). Measuring only eight bands minimizes the dimensions (32 × 32 × 42 cm) and mass (<10 kg) to those required for microsatellites. Although infrared spectroscopy methods have traditionally focused on Earth's atmospheric windows, including near-infrared (<5 µm; Bibring et al., 2006; Carrozzo et al., 2012; Viviano-Beck et al., 2014, 2017) and thermal infrared (8–14 µm; Fergason et al., 2006; Kubiak and Stach, 2013; Ciazela et al., 2021, 2023), interest in longer wavelengths stimulated by space applications has recently increased (Rogalski, 2017). However, the biggest challenge related to the designed device was the small field of view conditioned by the high resolution required by the purpose of the research (5 m/pixel), which in such a limited space can only be achieved using the Cassegrain optical system.

## 5    Simulated Mass Absorption Coefficients for Various Mineral Mixtures

Large monomineral fields on the Moon, Mars, or Earth are rare, especially for ore minerals. Ore minerals are usually dispersed in host rocks. The spectra emitted by rocks represent a mixture of ore (Figure 6A) and rock-forming minerals (Figure 6B). While the most common ore mineral on the Moon is probably ilmenite (Papike et al., 1976; Bhatt et al., 2012, 2015), and to a lesser extent, troilite and pyrite, basalts are the most common rocks found in the Apollo program and lunar meteorites if we will consider only lunar mare. To date, >400 meteorites have been classified as samples from the Moon (Korotev and Irving, 2021). Except for breccia typical for impacts, most represent lunar mare basalts, other lunar basalts, or anorthositic highlands rocks (Korotev and Irving, 2021).

Using the linear mixing method analogical to the used by Ciazela et al. (2022), we simulated various mixtures of ilmenite and a mixture of Ti-rich lunar basalt silicates (olivine, pyroxene, and plagioclase in a ratio of 5:50:30) to determine what would be the minimum amount of ilmenite needed to detect it from orbit (ratios of 1:9, 2:8, 3:7, and 4:6 of ilmenite to basalt were examined) (Figure 8A). Linear mixing of spectra is widely used in various scientific fields, from remote sensing to medicine and biology (Shi and Wang, 2014; De Angelis et al., 2017; Ciazela et al., 2022). Thermal IR spectra of multi-mineral surfaces can be modeled using linear combinations of the spectra of each mineral, weighted as percentages of the area occupied (Lyon, 1965; Thomson and Salisbury, 1993; Ramsey and Christensen, 1998). Comparing the results obtained by linear mixing and radiative transfer algorithms on Mars showed a good agreement between these algorithms (Christensen et al., 2000; Smith et al., 2000). Therefore, linear mixing can combine emissivities (Figure 8) or mass absorption coefficients (Figure 9) measured or entered from spectral libraries for single minerals into a surface spectrum. Here, the minerals and their proportions are selected based on the predicted composition of lunar rocks such as basalt and its variants (Papike et al., 1976). We used the C++-based PFSLook program Zalewska et al. (2016; 2019) described for linear emissivity mixing. In addition to simulations for ilmenites based on emissivity (Figure 8), we also performed simulations for pyrites based on mass absorption coefficients (Figure 9), for which we used our MATLAB-based code (see Supplementary Material 1).

However, to acknowledge the notion that linear mixing is only successful for particles much larger than the wavelength (Thomas and Salisbury, 1993), which is often the case for the Moon (~50% particles >100 μm) but it is not always the case (~50% particles <100 μm) (Ryu et al., 2018), we have performed additional single-scattering albedo/Hapke modeling (Harris and Grindrod, 2018) (Supplementary Figure 1) and compared the results between the two models. The position of the ilmenite peaks is slightly shifted between the two models, and this shows that laboratory test of the MIRORES instrument will need to include physical mixtures of the ore minerals of interest with lunar regolith of different grain sizes before selecting the final position of the detectors in the flight version.

Our simulations indicate that the proposed method can detect fields containing 10–20% pyrite or ilmenite (Figures 8 and 9). In MIRORES' most minor possible field of view of 3.3 m × 5.0 m = 16.5 m$^2$ for an orbital height of 50 km, pyrite or ilmenite can be detected from an orbit if they occupy 2–3 m$^2$. Such sensitivity should allow the detection of volcanogenic massive sulfides (VMS) and their stockworks. For example, the Rio Tinto VMS deposits in the Iberian Pyrite Belt contain two large ore deposits with massive pyrite, San Dionisio (~600 m × 130 m = 78,000 m$^2$) and Filon Sur (~1,100 m × 50 m = 55,000 m$^2$), but also more extensive pyrite stockworks around the Salomon and Argamasilla sites (640 m × 280 m = 179,200 m$^2$) determined to contain >20 wt% S (Martin-Izard et al., 2015). This amount of S corresponds to 23–37 vol.% pyrite considering that pyrite contains 53 wt% S and has a density of 5.0 g/cm$^3$, with other minerals having a 2.6–5.0 g/cm$^3$. Such a volume is much more than 10–20 vol.% needed for MIRORES detection. Earth analogs help predict where ore minerals should be searched on the Moon as similar ore-forming processes, such as volcanism and hydrothermal activity, occurred there. Data from lunar meteorites and the Apollo program indicate that the mineral composition on the Moon is similar to that on Earth. The same is the case for rocks whose proportions are different than on Earth, but similarly to our planet, the bottom of the oceans is mainly covered with basalts, and acidic rocks dominate the continents (highlands).

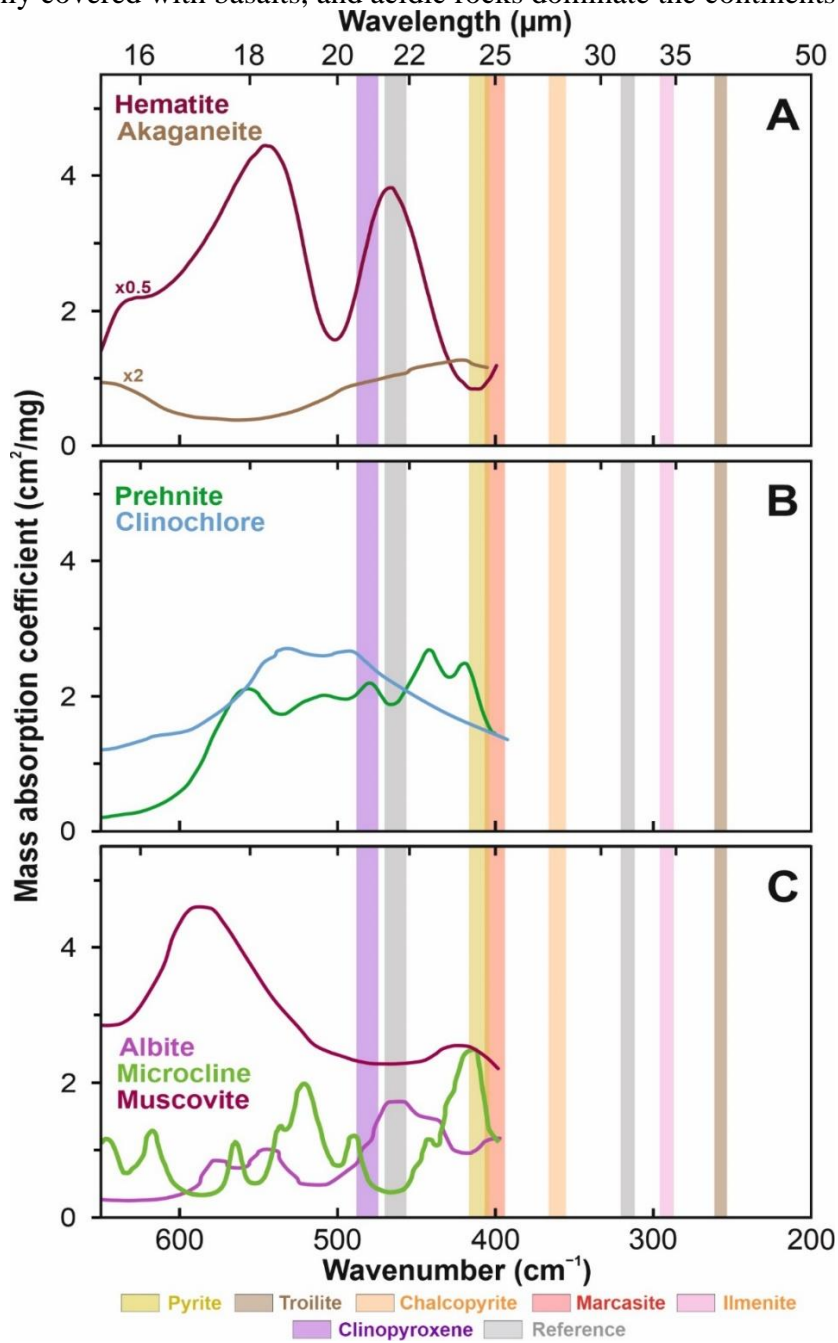

**FIGURE 7.** Mass absorption coefficients of critical lunar minerals with characteristics similar to sulfides and ilmenites grouped into **(A)** oxyhydroxides (Stimson and O'Donnell, 1952 and JHU spectral library: https://speclib.jpl.nasa.gov/documents/jhu_desc; Salisbury et al., 1989; Murad and Bishop, 2000; Kendix et al., 2008) with values multiplied by 0.5 (hematite) and 2 (akaganeite) for clarity, **(B)** secondary rock-forming silicates (Stimson and O'Donnell, 1952 and JHU spectral library: https://speclib.jpl.nasa.gov/documents/jhu_desc; Salisbury et al., 1987, 1989), and **(C)** rarer primary rock-forming silicates (Salisbury et al., 1987). The rectangles indicate the spectral ranges of the eight detectors: brown (254–262 cm$^{-1}$ = 38.22–39.38 µm: troilite), pink (288–297 cm$^{-1}$ = 33.69–34.71 µm: ilmenite), orange (357–368 cm$^{-1}$ = 27.19−28.01 µm: chalcopyrite), red (396–408 cm$^{-1}$ = 24.53−25.27 µm: marcasite), yellow (405–418 cm$^{-1}$ = 23.94–24.66 µm: pyrite), gray (313–322 cm$^{-1}$ = 31.03–31.97 µm and 458–472 cm$^{-1}$ = 21.18–21.82 µm: reference bands) and purple (476–490 cm$^{-1}$ = 20.39–21.01 µm: clinopyroxene).
14

# 6 Potential geological targets on thee Moon

## 6.1 Volcanic regions

Researchers focus mainly on areas of volcanism penetrated by hot hydrothermal waters. Post-magmatic or groundwater heating above magma chambers or igneous intrusions reaches several hundred degrees Celsius, enabled by high pressure. Hot water dissolves significant amounts of sulfur and metals. After rising to the surface and rapidly cooling, the metals and sulfur precipitate to form sulfide deposits. Water needed for hydrothermal processes is present on the Moon in a few areas, mainly as ground ice. Hot, geologically active regions do not exist, although volcanism was likely common in the first half of the Moon's history (Meyer, 2003; Li et al., 2021). Large-scale volcanism ceased, but the youngest volcanic provinces showed sustained volcanic activity until <100 Ma (Braden et al., 2014; Lapôtre et al., 2020). Therefore, relatively young hydrothermal circulation and associated ore deposits can be expected near volcanic cones, fissures, and outgassing structures (Kartashov et al., 2018).

While hydrothermal metal ore deposits on the Moon are possible, other more efficient ways exist to concentrate oxide and sulfide minerals into exploitable ores. Settling dense mineral crystals (for example, ilmenite, chromite, and platinum group minerals) is possible in silicate magma if the magma remains melted long enough. On Earth, such accumulations usually occur in layered intrusions. These bodies are formed from large masses of magma that have crystallized in the rocks of the Earth's crust without reaching the Earth's surface. Under these conditions, cooling is slow, and physical separation processes, including convection and mineral settling, have time to act (Papike et al., 1991). Well-known examples of ore deposits resulting from these processes occur at the Stillwater Complex in Montana (Raedeke and Mccallum, 1984) and the Bushveld Igneous Complex in South Africa (Mungall and Brenan, 2014).

Similar layered intrusions may form on the Moon. The accumulation of dense oxide and sulfide minerals as they crystallize from magma depends on their rate of descent into the less dense silicate melt. The higher the rate of descent, the harder the minerals can sink before the melt solidifies and the greater the chance of accumulating them. Stokes' law controls the rate of descent, being directly proportional to gravity and inversely proportional to viscosity. Gravitational acceleration on the lunar surface is 6 times lower than on Earth, but the viscosity of an average lunar basalt melt is 10 to 100 times lower than the viscosity of terrestrial basaltic magma. The reduced viscosity, therefore, outweighs the effect of the lower lunar gravity, so the oxide and sulfide crystals will have higher settling rates on the Moon. Therefore, there may be layered ore deposits on the Moon similar to those on Earth or even more significant than those on Earth (Papike et al., 1991). This process is one of the reasons why mafic rocks so rich in ilmenites are observed on the lunar surface.

## 6.2 Impact craters

Currently, hydrothermal deposits can form after meteorite impacts. Stripped of its protective atmosphere, the Moon is still bombarded by meteorites. After the impact, the rocks melt in the same way as in volcanic processes, and the groundwater in their vicinity behaves like hydrothermal waters, leading to the local formation of deposits on the rims of craters. On the Moon, all meteorite crater rims should be suitable for remote sensing of ore minerals due to potential impact-induced water-solid reactions (Stopar et al., 2018). Impact-induced heating of the volatile-enriched subsurface or its water vapor-induced alteration has been proposed as a result of observations of several post-impact crater basins, with near-surface temperatures >1000°C and/or more extended periods (thousands of years) of more moderate heating to ~300°C, to explain the temperature metamorphism observed in some Apollo rocks (McKay et al., 1972; Hudgins et al., 2008; Shearer et al., 2014). Even at lower temperatures, some minerals, such as olivine, react relatively quickly with unsaturated liquid water (Stopar et al., 2006). Chemical reactions are generally enhanced by surface defects such as cracks, impact defects, radiation damage, and increased amorphousness of structures (Holmes et al., 1973; Brantley et al., 1986; Petit et al., 1989; Hamilton et al., 2000; White and Brantley, 2003), all of which should be relatively common in the harsh lunar regolith and megaregolith. Due to near-surface water ice in some polar regions suggested by the Lunar Reconnaissance Orbiter (Spudis et al., 2013; Lucey et al., 2014; Hayne et al., 2015) and potential hydrothermal changes, future missions must include instrumentation to detect trace or minor hydrothermal changes in the geochemistry of minerals and rocks, as well as potential mineral transformations including iron oxides, hydroxides, sulfides, amorphous silica, and secondary clay minerals (Stopar et al., 2018).

In impact craters >10 km (Osinski et al., 2013), hydrothermal mineralization, such as lenses of massive pyrite, may be significant enough to be detected with MIRORES. Large-scale impact-related ore mineralization is known from several impact craters on Earth, such as the Sudbury Basin in Canada (Dare et al., 2010) or the Witwatersrand supergroup in South Africa (Frimmel, 1997). In the Sudbury Basin, sulfide ores composed of massive sulfides, pyrrhotite, and chalcopyrite, such as in the Fraser, McCreedy East, and Longvack areas, are up to several hundred meters long and could be detected by MIRORES (Li and Naldrett, 1994; Dare et al., 2011). The Rio Tinto VMS and the impact-related Sudbury mineralization contain one of Earth's highest-known concentrations of massive sulfides (Martin-Izard et al., 2015). However, even smaller mineralizations covering only a few square meters can be easily detected on the Moon with the MIRORES instrument thanks to its high resolution.

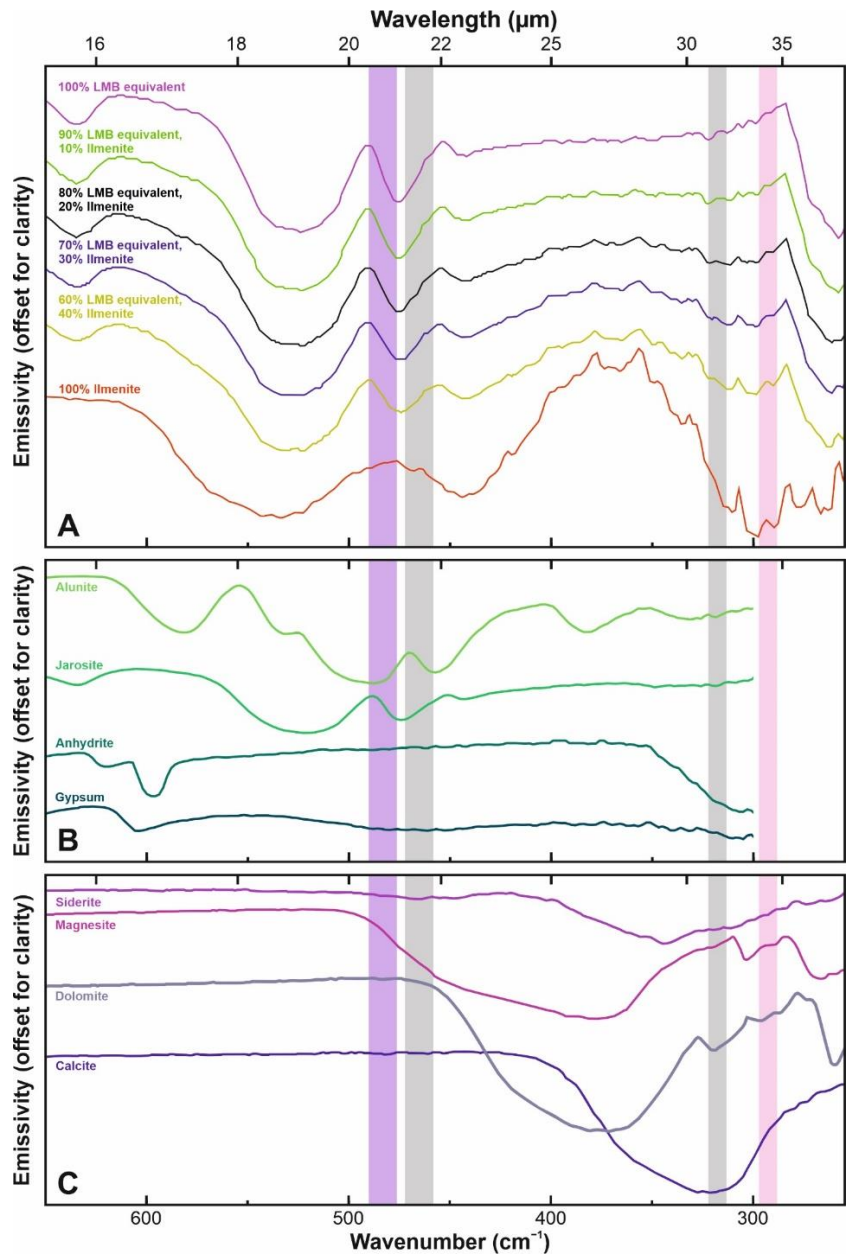

**FIGURE 8.** Panel **(A)** shows the emissivity of ilmenite from the Arizona State University (ASU) spectral library (https://speclib.asu.edu/, ilmenite #463), equivalents of lunar mare basalts (LMB), and their various mixtures. Pyrite data has been provided and discussed by Ciazela et al. (2022). To the best of our knowledge, other minerals are not measured for emmisivities due to a complex preparation procedure requiring temperature control (Ciazela et al., 2022). Therefore, they will still need to be measured in the future, but to understand their spectral features' potential position and intensity, we discussed their mass absorption coefficients in Figures 6 and 7. Here, the composition of the lunar mare basalt equivalents is the same as in the Section 5 text but normalized from 85% to 100%. The input spectra of olivine (fayalite), plagioclase (labradorite), and pyroxene (pigeonite) are also from the ASU spectral library, as are the sulfate and carbonate spectra collected in panels **(B)** and **(C)**. Various ilmenite and lunar mare basalt equivalent mixtures were simulated using the PFSLook software (Zalewska et al., 2019). The spectral ranges of four of the eight detectors are marked with rectangles: pink (288–297 $cm^{-1}$ = 33.69–34.71 μm: ilmenite), gray (313–322 $cm^{-1}$ = 31.03–31.97 μm and 458–472 $cm^{-1}$ = 21.18–21.82 μm: reference bands), and purple (476–490 $cm^{-1}$ = 20.39–21.01 μm: clinopyroxene). The ordinate axes do not have a numeric scale because the graphs are shifted for clarity.

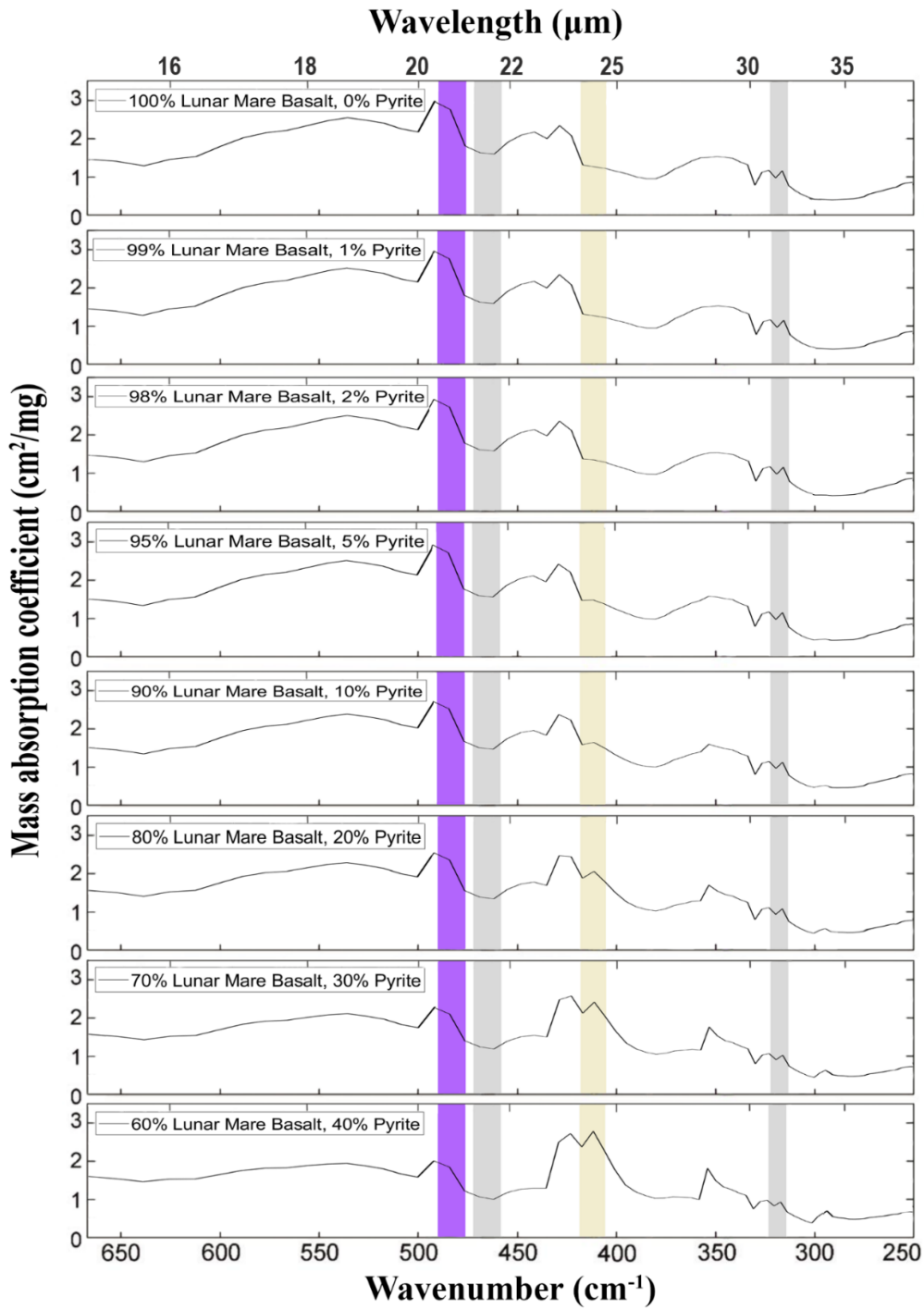

**FIGURE 9.** Mass absorption coefficients simulated for surfaces composed of various proportions of pyrite and lunar mare basalt equivalent. In this simulation, the composition of lunar mare basalts was assumed for basalts with a moderate content of $TiO_2$, composed of silicates with proportions of 60 pyroxene: 30 plagioclase: 5 olivine. The input spectra are taken from Figure 6. The spectral ranges of four of the eight detectors are marked with rectangles: yellow (405–418 cm$^{-1}$ = 23.94–24.66 μm: pyrite), gray (313–322 cm$^{-1}$ = 31.03–31.97 μm and 458–472 cm$^{-1}$ = 21.18–21.82 μm: reference bands), and purple (476–490 cm$^{-1}$ = 20.39–21.01 μm: clinopyroxene). There is a partial interference between the pyrite and the clinopyroxene secondary peak on the right (at ~421 cm$^{-1}$ = 23.8 μm, cf. Figure 6), which becomes significant when the pyrite is <5 wt.%. Equation 1 eliminates the signal from this interference. The mineral mixtures were calculated using the MATLAB script in Supplementary Material 1.

# 7      Goals of the lunar MIRORES mission and its justification

Using the MIRORES spectrometer will serve two primary and four secondary scientific goals. The primary goals are 1) the selection of basalt mare areas on the Moon with the most significant resource potential and 2) the indication of strategic places for establishing human bases on the Moon based on the acquired data and other geological parameters. The critical resources given the resource potential are a) metals, including Fe and Ti associated with oxide mineralization, especially ilmenite, as well as Cu, Ni, and PGE associated with sulfide mineralization. These two groups of minerals usually coexist (Hu et al., 2021; Saal and Hauri, 2021)]; b) sulfur associated with sulfides and possibly sulfates; and c) helium-3, helium-4, and other trace gases captured from the solar wind by ilmenite (Johnson et al., 1999; Fa and Jin, 2007; Kim et al., 2019).

The secondary goals are 1) demonstration of mineralogical differences in the ore mineralization between the surfaces of the lunar mare and the anorthositic highlands, 2) verification of differences in ore mineralization between the lunar circumpolar areas and other areas, 3) providing new insights into the geology of KREEP areas enriched in K, REE, and P, located on the basalt mare's margins (Jolliff et al., 2000), which can be characterized for ore mineral content but also apatite (robust spectral features between 25–30 μm), as well as 4) identifying the potential interstellar object (ISO) craters on the Moon's surface (Cabot and Laughlin, 2022) if those are significantly enriched in metals.

The primary goals result from the need to prepare to construct lunar bases. This construction should begin with the Artemis program's development, especially the first human-crewed flights starting with the Artemis 3 mission in 2025. Advantageous is the fact that many different metals usually coexist and are concentrated in sulfides and oxides, often constituting the same paragenesis (Vaniman et al., 1992; Hu et al., 2021; Saal and Hauri, 2021). Therefore, the content of different metals is usually elevated in the same areas. The highest concentrations of both $TiO_2$ (~10 wt%) and $FeO_T$ (~23 wt%) were recorded in the same basins: Oceanus Procellarum and Mare Tranquillitatis, and to a lesser extent in Mare Imbrium (cf. Figure 3 and Lucey et al., 1998). However, the concentrations of rarer metals occurring as accessory or trace elements, especially Cu, Ni, and PGE, have never been measured from orbit. Since these elements are concentrated primarily in sulfides (Goldschmidt, 1937; Lee, 2018), sulfide detection from orbit will be crucial for geological exploration. MIRORES has such abilities. Sulfur is also concentrated mainly in sulfides due to the scarcity of water-loving sulfates on the Moon and has not been the subject of orbital research so far. The last resource group, volatiles, not in the lattice of ore minerals, are captured from the solar wind by ilmenite (Johnson et al., 1999; Fa and Jin, 2007; Kim et al., 2019). Hence an excellent correlation between Ti and He content in the Apollo samples (Figure 1) (Johnson et al., 1999; Fa and Jin, 2007). Like He, other gases, including $O_2$, $H_2O$, or OH, are also probably trapped by ilmenite (Kim et al., 2019).

The acquired mineralogical data will allow the selection of strategic places for human lunar bases when combined with other geological parameters such as water ice content, geomorphology, and daily temperature changes. In the long run, importing materials for constructing and developing lunar bases from Earth will be unsustainable. It will be cheaper to produce the necessary materials on-site, which will include sulfur concrete (Grugel and Toutanji, 2008), oxygen-based rocket fuel (Lomax et al., 2020), pyrite-based solar panels (Kristmann et al., 2022). In addition, there is great interest in extracting $^3$He from ilmenite (Johnson et al., 1999; Fa and Jin, 2007; Kim et al., 2019) to transport it to the Earth, enabled by the high market price of this isotope. It will fuel future thermonuclear fusion power plants, such as ITER in France. Helium-3 is an alternative to tritium, which causes the emission of many neutrons, seriously complicating the work of future thermonuclear fusion reactors.

The scientific needs determining the secondary goals should also be emphasized. First, we must demonstrate the ore mineral differences between the lunar mare basalts and the anorthositic highlands, which is already possible in the mission's first stage. The studies carried out so far, mainly by the missions Clementine (1994) and Lunar Prospector (1998–1999), indicate geochemical differences between the two main types of rocks on the Moon, and especially significantly higher Fe content (on average 18 wt% $FeO_T$, although in some areas 22–23 wt% $FeO_T$) and titanium (up to 13 wt% $TiO_2$) in lunar mare basalts compared to anorthosites (3.5–5.5 wt% $FeO_T$ and <0.5 wt% $TiO_2$, respectively) (Lucey et al., 1998; Giguere et al., 2000; Lawrence et al., 2002). Later studies of the $M^3$ spectrometer onboard Chandrayaan-1 confirmed mineralogical differences between these two groups of rocks, but only in rock-forming minerals. First, lunar mare basalts show 35–50 vol.% plagioclase (Bernhardt et al., 2017; Corley et al., 2018), and lunar highland anorthosites >90 vol.% plagioclase (Cheek et al., 2013; Bernhardt et al., 2017). However, no maps show the differences between the domains of basalt mare and anorthositic highlands regarding their content of ore minerals, such as ilmenite, troilite, and others. The only mineralogical data on ore minerals come from site-focused analyses of the Apollo missions (Papike et al., 1991) and the Chinese Change-5 mission (Hu et al., 2021), which do not allow mapping.

The second secondary goal is verifying differences in ore mineralization between the lunar circumpolar and other areas. Since landers have not explored the polar regions, the only comparative data comes from orbital observations regarding chemical and mineral composition. These areas show slightly elevated $TiO_2$ (1–5 wt%) and $FeO_T$ (5–15 wt%) relative to the anorthositic highlands (Lucey et al., 1998), which would imply an increased ore mineral content. The orbit of the Chandrayaan-1 mission did not allow the collection of mineralogical information from these areas by the $M^3$ spectrometer (Bernhardt et al., 2017). Therefore, the role of MIRORES in recognizing polar regions may be crucial.

The third secondary goal will be insights into the geology of KREEP sites, which are located at the margins of lunar mare (Jolliff et al., 2000) and can be characterized by MIRORES for ore minerals, but also apatite (robust spectral features between 25–30 µm) rich in REE and containing water (Liu et al., 2022). So far, we have chemical data (Lucey et al., 1998) and rock-forming mineral data (Bernhardt et al., 2017) that indicate values intermediate between those for the lunar mare and the anorthositic highlands (Lucey et al., 1998; Bernhardt et al., 2017). However, orbital data on ore mineral and apatite contents are missing. Apatite and ore minerals may deviate from the general pattern obtained from bulk-rock geochemistry and rock-forming minerals.

The discoveries of 'Oumuamua (from the Pan-STARRS survey; Meech et al., 2017) and comet 2I/Borisov (Guzik et al., 2020) have prompted intensive study of the number density, composition, and origin of ISOs. Motivated by an encouragingly high encounter rate of ISOs, up to about seven per year, that pass within 1 AU of the Sun, Cabot and Laughlin (2022) have proposed an alternative route to characterize ISOs by identifying their impact craters on terrestrial solar system bodies. For example, the molten and vaporized projectile matter may mix with impact-modified target rock (impactites) and impart telltale chemical signatures. More optimistically, some projectile material might survive in a solid phase. A standard chemical and isotopic analysis exist for characterizing meteorites and impact melts (Tagle and Hecht, 2006; Joy et al., 2016), which could reveal the ISO's composition.

# 8      MIRORES tasks on the orbit

The task of the MIRORES mission will be to make high-resolution maps of the content of ilmenite, other potential oxides, and significant sulfides in orbits with the largest possible inclinations (> 60°, optimally close to 90°) passing through the areas of basalt mare with the highest Ti and Fe contents (Figure 10): 1) Mare Tranquillitatis (26–31°E → 150 km wide at the equator), 2) Oceanus Procellarum (53–57°W → 120 km wide at the equator), and 3) Mare Imbrium (16–24°W → 240 km wide at the equator). This work will be carried out in two stages, general (years 1–3) and detailed (year 4).

The first stage of the mission, lasting up to three years, will allow the creation of lower-resolution maps of oxide and sulfide distribution, which will be used to determine the most promising areas to be analyzed in detail in the second stage. In the first stage, a precessing polar orbit with an altitude of

~200 km and an orbital period of ~2 h will be used. With the assumed very small precession of the suborbital point on average between 20–40 m at the equator (such precession allows the field of view to cover all of the areas mentioned above), a 240–480 m wide belt along the equator can be covered with observations per day. One of the areas (a), (b), or (c) can be fully or almost entirely covered during the year. Therefore, by changing the orbit approximately once per year, all three areas can be covered by observations (Figure 10).

These are the optimal assumptions of the mission, requiring relatively large fuel resources and the experience of the Launch Service Provider (LSP) company, which will allow the selection of optimal orbits. With a more modest mission budget, selecting only one or two areas is considered, from which priority would be given first to Mare Tranquillitatis and secondly to Oceanus Procellarum. In addition, near-equatorial orbits with low inclinations could be considered instead of polar orbits to avoid counteracting accelerations related to lunar rotation.

This stage will result in mineralogical distribution maps of ilmenite and other potential oxides such as hematite and magnetite, as well as troilite and other sulfides. High-resolution maps (~30 m/pixel) of the occurrence of these minerals will surpass the accuracy of chemical maps of Ti and Fe distribution with a resolution of ~300 m/pixel by order of magnitude, and for the first time in the history of lunar exploration, will provide mineralogical data on ore minerals, more critical from the ore deposit point of view than chemical data.

In the second stage, the maps obtained during the first exploration stage will allow the selection of the most promising areas for preparing detailed maps with even higher resolution (<5 m). These areas will be mapped by changing the altitude from 200 km to 50 km. Such a strategy will allow the selection of three to ten key areas for the detailed study of the second stage, considering the size of specific deposit areas on Earth (Li and Naldrett, 1994; Dare et al., 2011; Martin-Izard et al., 2015).

Although shorter, the second leg of the mission will involve relatively higher fuel consumption due to the necessary orbital maneuvers to keep the right orbit despite the regional and local mass concentrations (called mascons) observed on the lunar surface. They cause a lowering of the orbit, especially when it passes low over the surface. Counteracting this phenomenon causes an additional increase in fuel consumption (Konopliv et al., 2001). The fuel consumption should be estimated in the subsequent stages of mission planning. If the estimated fuel consumption to maintain a 50 km periselenium orbit exceeds mission capabilities, raising the periselenium orbit to 70 km or even 100 km should be considered.

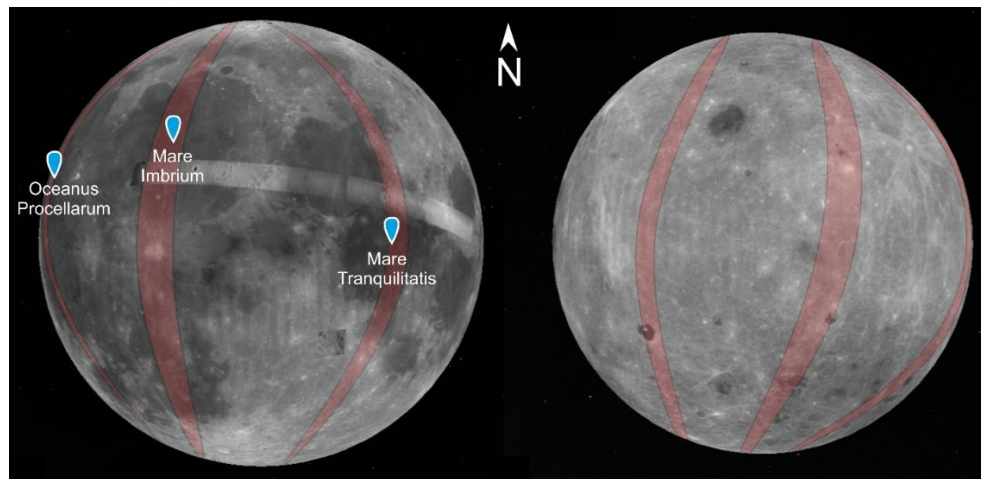

**FIGURE 10.** Examples of orbits selected for studies with the MIRORES instrument assuming an inclination of 90°. The orbits pass over the mare selected for research: Oceanus Procellarum, Mare Imbrum, and Mare Tranquillitatis. Alternatively, near-equatorial orbits passing through all three basins could be considered to limit fuel consumption. The Moon's near side with basalt mare is on the left, and the far side is on the right.

# 9 MIRORES design for a lunar mission

The lunar version of the MIRORES instrument (Figures 11–13) is developed from its Martian version (Ciazela et al., 2022), and all the technical descriptions not given here are the same as those reported by Ciazela et al. (2022). The critical difference is that thanks to the lower $H_2O$ and $CO_2$ contents in the lunar atmosphere, we can measure a broader spectral range of the FIR spectrum extending from 20 to 40 µm. Here, prominent absorption peaks are displayed by most sulfides and oxides, including not only proposed previously (Ciazela et al., 2022) chalcopyrite, marcasite, and pyrite, which is ubiquitous in all planetary bodies but also ilmenite with the most vital line at 34.2 µm and troilite with the most vital line at 38.8 µm (Figure 6). In addition, we will measure two bands at 21.5 µm and 31.5 µm for reference (grey bands in Figure 6), with the latter differing from the Martian version (26.5 µm, Ciazela et al., 2022). The lunar version will also differ with the position clinopyroxene detector, measuring at 20.7 µm (purple band in Figure 6) in this case, which will make it possible to remove partial interference on pyrite from the secondary peak of clinopyroxene at ~23.8 µm using the following formula (Eq. 1):

$$Py_{24.3_{Cpx}} = \frac{Cpx_{20.7} - R_{31.5}}{25} \qquad (1)$$

where $Py_{24.3_{Cpx}}$ is the contribution of the secondary peak of clinopyroxene to the absorption peak at the pyrite (Py) detector, $Cpx_{20.7}$ is the signal from the detector monitoring the intensity of the clinopyroxene (Cpx) signal, and $R_{31.5}$ is the signal from the reference detector at 31.5 µm. Here, 25 is the ratio between the absorption peak amplitude at the peak of clinopyroxene at 20.7 µm and the absorption peak amplitude of a part of the clinopyroxene ~23.8 µm peak within the pyrite detector. As for the Martian version (Ciazela et al., 2022), the equation is free of any assumption on the modal content of clinopyroxene on the surface. We only operate on the ratio between the two absorption features of clinopyroxene, which should be relatively constant independent of the content.

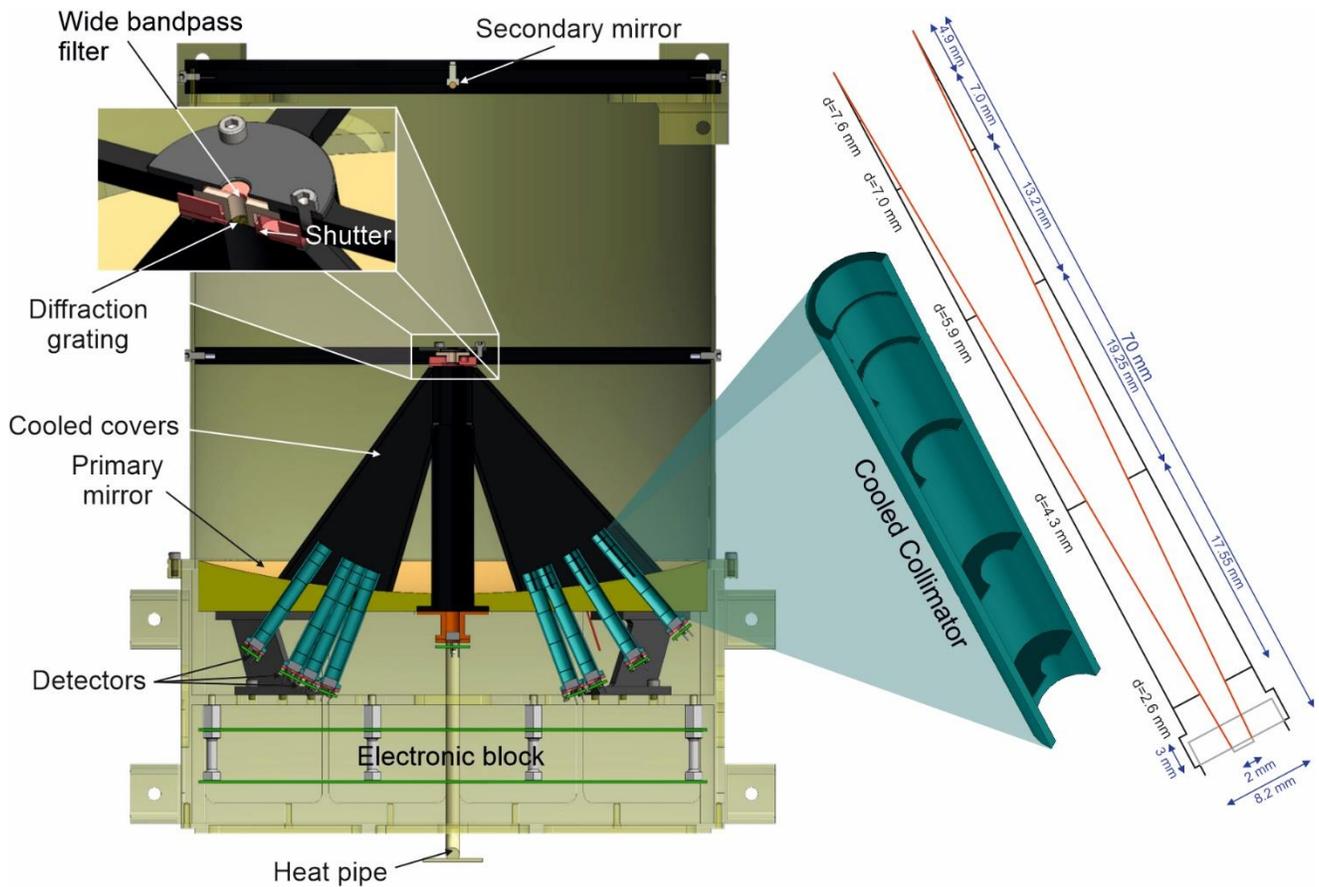

**FIGURE 11.** Simplified cross-section of the lunar MIRORES instrument. The size of the instrument is 42 × 32 × 32 cm. The eight main detectors are placed behind the eight turquoise collimators, the optical design of which is depicted on the right. The zoom shows the shutter's position in front of the diffraction grating. The wide bandpass filter in front of the shutter transmits radiation in the 20–40 μm range. The design is based on the Martian version of the instrument, but the detector and collimator (Ciazela et al., 2022) set up is substantially different to include ilmenite and troilite typical for lunar ores.

For the lunar version, the primary mirror, made of Au-coated alumina 6061 covered with a protective $SiO_2$ layer, with a diameter of 300 mm, reflects the radiation into the smaller secondary mirror made of Au-coated stainless steel covered with a protective $SiO_2$ layer with a diameter of 4.1 mm. Note that the longer the wavelength, the smaller the difference between the reflectance of Fe, Al, Au, and Ag, and it becomes nearly negligible >5 μm. Alumina reflectance is >99% for > 20 μm, and Fe reflectance is 98% for >12 μm (Bass et al., 1995). Wroclaw Technology Park (Wrocław, Poland) will manufacture primary and secondary mirrors. The distance between the primary and secondary mirrors is 300 mm (Figure 12). The telescope's effective focal length is 30.3 m, and the detector window diameter is 2 mm. For the lowest planned altitude of 50 km, the field of view in a static position is thus 0.002 m through 30.3 m times 50,000 m, which makes 3.3 m. Given that the satellite will move during the 1 ms measurement with an orbital speed of 1656 m/s, the shift related to the position change is 1.7 m. Therefore, the length of the field of view along the satellite path will be 5 m, so the field of view can be reported as 3.3 m × 5.0 m (16.5 m$^2$).

All radiation reflected from the secondary mirror is directed to the wide bandpass filter (20−40 μm), followed by the shutter and diffraction grating (Figure 11). The diffraction grating in the lunar version is made of 30-μm-thick copper wire and an aluminum frame with grooves manufactured by Fluence Technology Ultrafast Laser Application Laboratory (Wrocław, Poland) using Jasper X0 femtosecond laser and has a pitch of 60 μm. These parameters allow us to obtain a spectral resolution between 0.6–1.2 µm. For example, the spectral resolution for the 20.7 µm detector is 0.62 µm, calculated as 20.7 µm through 33.3 (number of active slits, 16.7 active slits/1 mm).

Here, the broadband beam is split into eight refracted quasi-monochromatic beams that fall into the set of eight pyroelectric detectors (five for target minerals, two for reference, and one for clinopyroxene interference). The positions of the five primary and three reference detectors (Figure 12) are calculated from the refraction angle (θ) and the distance between detectors and a diffraction grating (200 mm), as explained in Ciazela et al. (2022). These positions are +114.00 mm (ilmenite at 34.2 µm–$Ilm_{34.2}$), +92.01 mm (chalcopyrite at 27.6 µm–$Ccp_{27.6}$), +83.00 mm (marcasite at 24.9 µm–$Mrc_{24.9}$), +71.64 mm ($R_{21.5}$), −68.99 mm ($Cpx_{20.7}$), −81.00 mm ($Py_{24.3}$), −105.01 mm ($R_{31.5}$), and −129.33 mm (troilite at 38.8 µm−$Tro_{38.8}$) from the optical axis (Figure 12). All detectors register radiation from the same area of the Moon. The non-refracted radiation portion travels to the ninth detector at 0.00 mm (Figure 12). This detector registers the broadband non-refracted radiation between 20–40 µm to monitor the brightness of the lunar surface.

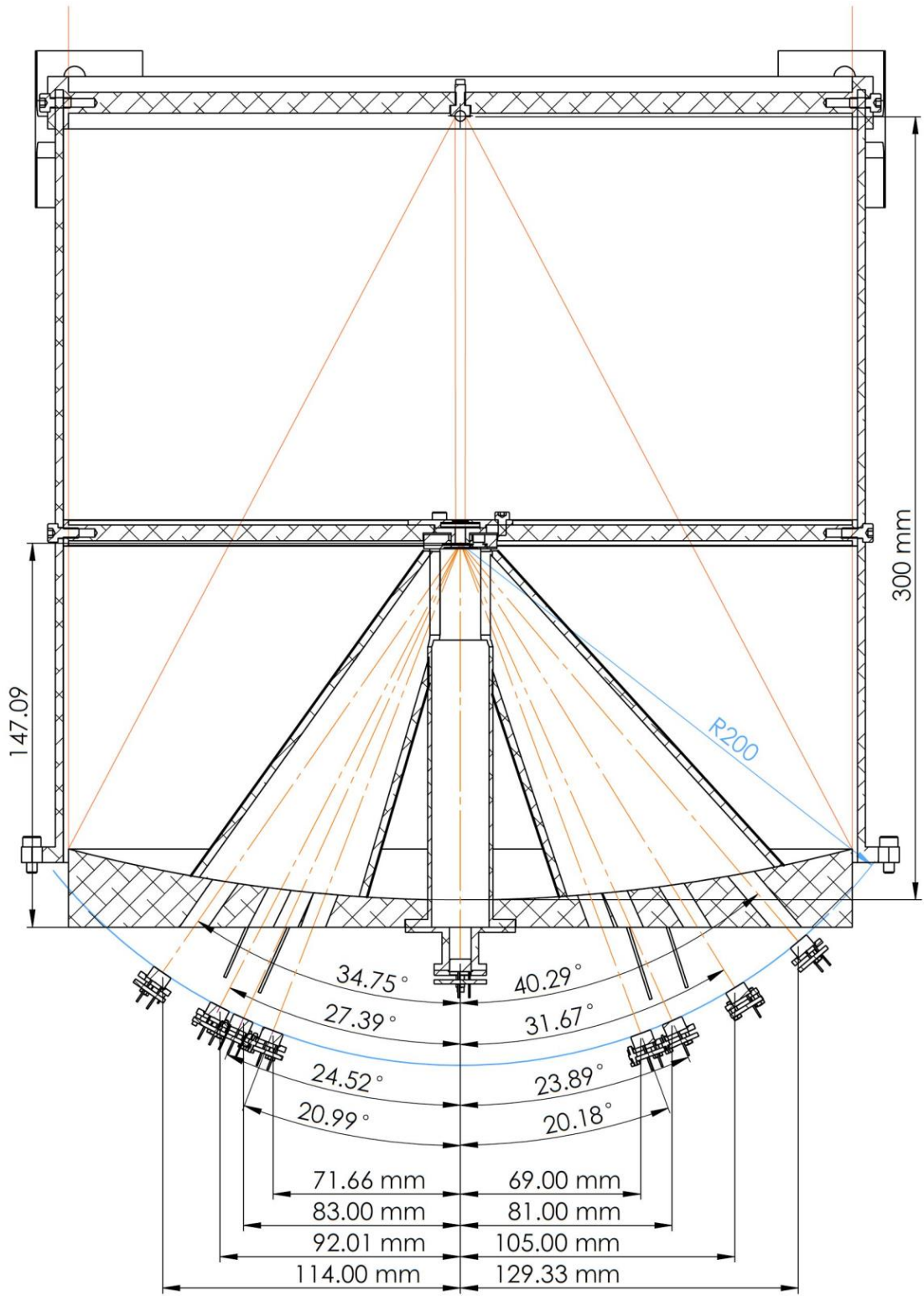

**FIGURE 12.** The optical system of the lunar MIRORES instrument with the primary mirror at the bottom and the secondary mirror at the top. Orange solid lines: incident radiation; orange dash-dotted lines: refracted radiation. R200 is the distance between the detectors and diffraction grating (200 mm). The thin blue curve marks this distance of 200 mm from the diffraction grating, and all the detectors are situated along this theoretical curve. The design is based on the Martian version of the instrument (Ciazela et al., 2022), but the detector setup is substantially different to include ilmenite and troilite typical for lunar ores.

As pyroelectric detectors are sensitive only to modulated radiation, we mounted a shutter ES6B (manufactured by Uniblitz, Rochester, NY, U.S.) before the diffraction grating and the wide bandpass filter (Figure 11). The detectors, described by Ciazela et al. (2022), are mounted on a copper plate to facilitate heat dissipation. Electronic circuits should have a radiation resistance parameter of 100 krad. Suitable are, for example, Microchip's RTAX FPGAs. Satellite shielding of the instrument is also necessary, optimally with a 10-mm-thick Al layer. The orbit on which MIRORES will be placed passes through areas of intense corpuscular radiation associated with coronal ejections of the Sun. This radiation consists of charged particles with high energy (mainly protons) that can damage the satellite's electronic components. Moreover, reaching the lunar orbit may require passing through the Van Allen radiation belts. Therefore, the electronics used in such an orbit must have adequate durability. The literature suggests that the arrival to the lunar orbit alone will be associated with the reception of ~1.2 krad by electronics (Märki, 2020). To calculate the maximum radiation dose, Total Ionizing Dose (TID), to which the device may be exposed, we also used the Spenvis software, which suggests that an unshielded device staying ten days in an average orbit of 38,400 km would take a TID dose >1,000 krad, but shielding of 10-mm-thick Al limits it to 13 rads. Simulations for the in-orbit instrument indicate that a two-year mission between 01/01/2026 and 01/01/2028 would take over 300 krad due to solar protons, but with 10-mm-thick Al shielding, it would only take 1.7 krad.

The critical components of the instrument, the primary and secondary mirrors, will be connected to the satellite through heat pipes. Assuming that the mirrors are shielded from direct solar radiation by the satellite body and the opening of the tube is additionally protected by a wide bandpass filter (20–40 µm), they receive from 8 W (Moon's blackbody radiation) to 217 W (blackbody radiation + reflected radiation from the Sun). In the assumed model, the Moon has a minimum temperature of 140 K at night, 400K at day at the equator, the solar flux is 1373 W/m$^2$, the satellite temperature is 273 K, the space temperature is 4 K, and the electronic power dissipates up to 10 W. However, given that: 1) the temperature on the Moon is close to 140 K for half of the time and between 140–400 K for the other half, 2) the blackbody radiation grows geometrically, and not proportionally, with temperature, and 3) and the daily maximum temperature is 400 K only at the equator dropping <300K in high latitudes and <200 K at the poles, we estimated that on average the heat pipes need to expel only ~30–40 W to dissipate to be able to dissipate the entire heat. If extreme conditions are avoided for extended periods, predicted mechanical deformation due to estimated temperature changes should not significantly affect the quality of the optical system (Figure 13).

For radiance calibration, as in the case of the martian MIRORES instrument (Ciazela et al., 2022), we will require the satellite to enable observation of two sources at known temperatures, deep space and an outer plate mounted on the satellite. Deep space is characterized by a constant temperature and should be measured at least twice per orbit. The plate should be measured every 10 min as it is expected to have variable but known temperature. The temperature can be measured continuously with an expected accuracy of 0.1 K.

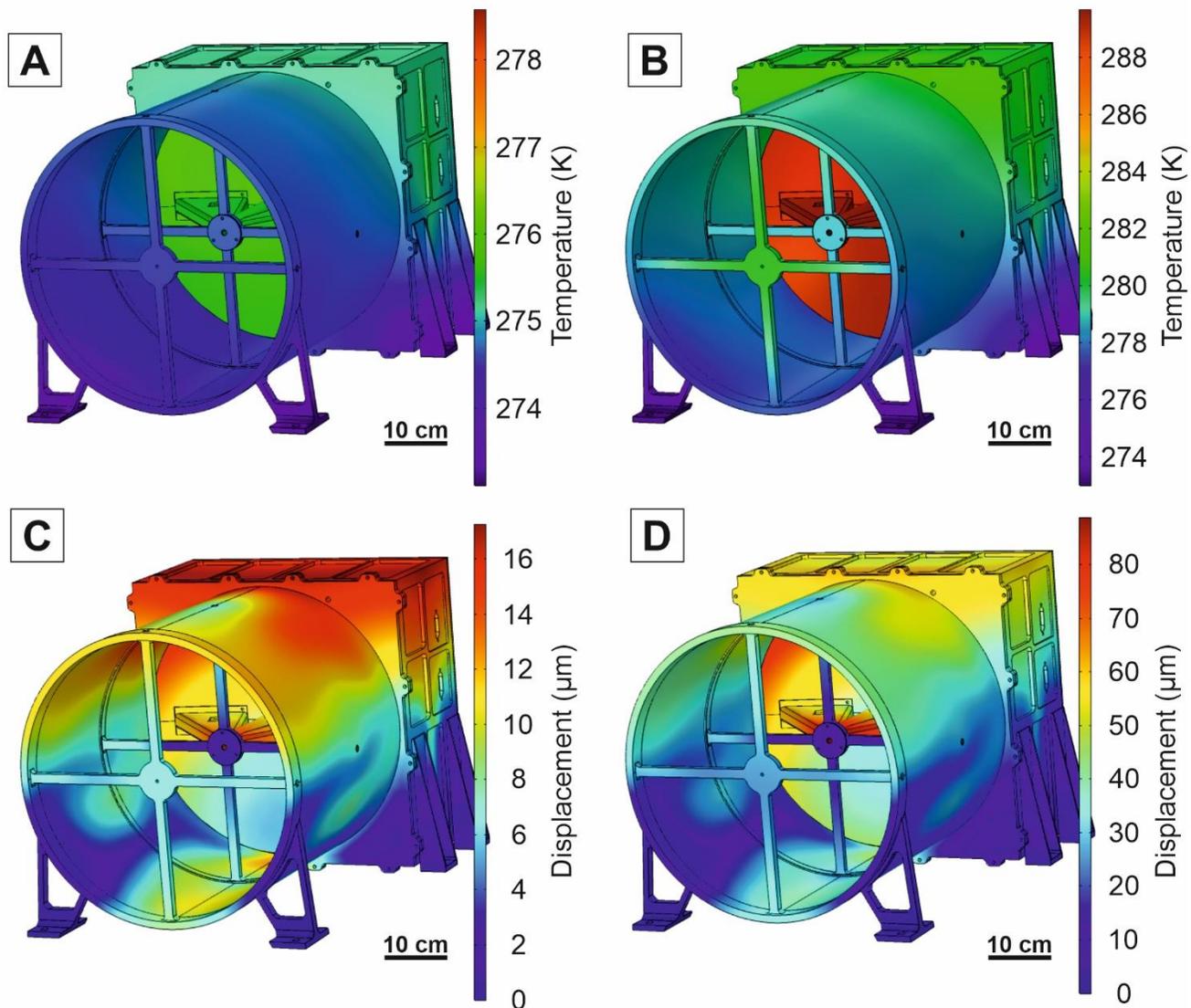

**FIGURE 13.** Thermal (A, B) and thermo-mechanical (C, D) analysis showing how changes in the thermal environment induce mechanical displacement that can affect optical performances. The simulation assumes a large longwave pass (<20 μm) filter installed at the aperture of the tube to reduce the amount of heat that comes to the primary mirror and electronics. The assumed heat sources are then limited to 8 W (A) in case of the night (Moon's blackbody radiation) and 217 W (B) in case of day (blackbody radiation + reflected radiation from the Sun; assumed albedo of 0.12). The instrument is assumed to operate in nadir geometry and to be shielded from direct solar radiation by the satellite shield. In simulations B and D, it was assumed that only the electronics box has a fixed temperature for the satellite of 273 K. In the remaining elements, the heat flux is dissipated by thermal conductivity. Although the maximum displacement is high for the collimator cover and the outer part of the primary mirror (without detectors), the displacement in the crucial part with the detectors is moderate.

## 10   Conclusions

Several ore minerals common on the Moon, including ilmenite and troilite, have prominent absorption peaks in the FIR wavelength range of 20–40 μm. The spectral features of sulfides and oxides in the FIR range are much stronger than those of any of the common lunar minerals, including olivines, clinopyroxenes, orthopyroxenes, plagioclase, major sulfates, and major carbonates. Our simulations indicated that fields containing 10–20% ilmenite or sulfide could be detected from orbit in the FIR range. However, although known from Martian instruments, Thermal Emission Spectrometer onboard

Mars Global Surveyor and Planetary Fourier Spectrometer on Mars Express, FIR is not yet applied on lunar orbiters.

Therefore, we designed a new instrument with a resolution that allows the detection of ore deposits. The use of the Cassegrain optical system achieved this. The field of view of 3.3 × 5.0 m achieved from an altitude of 50 km makes it possible to detect areas covered by 2–3 m$^2$ of ilmenite or pyrite on a surface of ~17 m$^2$, creating possibilities for detecting large and local small orebodies along with their stockworks. The instrument should be integrated into a satellite launched to the Moon by the Polish Space Agency between 2028–2029. Prototype construction for Earth-based tests started in 2021, and its laboratory testing began in 2022. The Earth test of the prototype from the drone will start in the early summer of 2023.

US-led efforts leading to the expected lunar landing within the Artemis program focus mainly on the polar areas (South Pole), and water ice is the primary resource of interest. The MIRORES mission proposal, on the other hand, focuses on the lunar maria. It is compatible with the Artemis program but does not directly overlap with the American and other proposals (Creech et al., 2022). It instead examines the alternative locations and promotes alternative ways of oxygen extraction (as a by-product of ore minerals processing rather than direct volatile prospecting). In this way, it constitutes a valuable supplement to the international lunar effort.

From a broader perspective, MIRORES may stimulate further efforts in developing lunar mining. Lunar resource mining investments can be divided into three stages. Stage 1 includes investment into enabling technologies of only high cost and is the current stage of development, with transport, non-commercial surveying, and preparation of extraction and processing technologies. Stage 2 will include commercial investment into mining data with moderate cost and return. In this stage, MIRORES may be adopted in commercial solutions. The mining data can be subject to commercial turnover before the mining equipment is integrated and sent to the surface. Stage 3 should include commercial investment into mining and ISRU activities with very high costs and returns. This stage will be developed when commercial mining activities become operational. Because Stage 1 is pending, and substantial public and private investments are allocated, the present paper is a part of preparatory activities for Stage 2. Since the subject activities are technically feasible, they also open the way for verifying commercial feasibility.

**Data Availability Statement**

The original contributions presented in the study are included in the article/Supplementary Material, further inquiries can be directed to the corresponding author. The dataset and Matlab script that generate data from Supplementary Figure S1 are openly available in figshare at https://dx.doi.org/10.6084/m9.figshare.25055618.

**Author Contributions**

Conceptualization, J.C., J.B., M.K., G.P., M.S., M.C.; Investigation, N.Z., M.C., G.P., J.B., M.K., M.S., J.C.; Methodology, J.C., M.K., L.S., N.Z., G.P.; Project administration, J.C., A.S., M.J., M.K.; Resources, N.Z.; J.B., M.K., Software, N.Z., M.K., J.B., J.C., G.P.; Supervision, J.C.; Visualization, J.B., M.K., N.Z., B.P., M.C., J.C., G.P., D.M.; Writing — original draft, J.C.; Writing — review & editing, J.C., M.J., M.S., M.K., J.B., B.P., M.C., A.S., M.R., N.Z., Z.S., D.M., and M.F. All authors have read and agreed to the published version of the manuscript.

**Funding**

The work was funded by the Polish Space Agency (POLSA) agreement no. PAK/U/167/2022/DBI/MBC, the European Space Agency (ESA) Space Resources project no. ESA AO/1-10824/21/NL/RA, and the National Science Centre Poland project no. 2020/37/B/ST10/01420 to J. Ciazela. L. A. Sterczewski acknowledges funding from the European Union's Horizon 2020 research and innovation programme under the Marie Skłodowska-Curie grant agreement No 101027721.

**Acknowledgments**

We thank S. Płocieniak for his advice on designing the primary and secondary mirrors, Fluence Technology (Warsaw, Poland, https://fluence.technology/) for the help in manufacturing the diffraction grating, as well as D. Jankowski, I. Badura, P. Boroń, A. Zwierzyński, and M. Laban for inspiring discussions. B. Pieterek thanks the Adam Mickiewicz University Foundation for the financial support in the academic year 2022/2023.

**Conflict of Interest**

The authors declare that the research was conducted in the absence of any commercial or financial relationships that could be construed as a potential conflict of interest.

**Supplementary Material**

Supplementary_material_1.m

Supplementary Figure 1

Properties of KLS-1 Lunar Simulant. *J. Aerosp. Eng.* 31. doi: 10.1061/(asce)as.1943-5525.0000798.

Saal, A. E., and Hauri, E. H. (2021). Large sulfur isotope fractionation in lunar volcanic glasses reveals the magmatic differentiation and degassing of the Moon. *Sci. Adv.* 7, eabee4641. doi: 10.1126/sciadv.abe4641.

Safari, M., Maghsoudi, A., and Pour, A. B. (2018). Application of Landsat-8 and ASTER satellite remote sensing data for porphyry copper exploration: a case study from Shahr-e-Babak, Kerman, south of Iran. *Geocarto Int.* 33, 1186–1201. doi: 10.1080/10106049.2017.1334834.

Salisbury, J. W., Walter, L. S., and Vergo, N. (1987). Mid-infrared (2.1-25 um) spectra of minerals: First edition.

Salisbury, J. W., Walter, L. S., and Vergo, N. (1989). Availability of a library of infrared (2.1-25.0 µm) mineral spectra. *Am. Mineral.* 74, 938–939.

Schlüter, L., Cowley, A., Pennec, Y., and Roux, M. (2021). Gas purification for oxygen extraction from lunar regolith. *Acta Astronaut.* 179, 371–381. doi: 10.1016/j.actaastro.2020.11.014.

Schwandt, C., Hamilton, J. A., Fray, D. J., and Crawford, I. A. (2012). The production of oxygen and metal from lunar regolith. *Planet. Space Sci.* 74, 49–56. doi: 10.1016/j.pss.2012.06.011.

Sekandari, M., Masoumi, I., Pour, A. B., Muslim, A., Rahmani, O., Hashim, M., et al. (2020). Application of Landsat-8, Sentinel-2, ASTER, and WorldView-3 Spectral Imagery for Exploration of Carbonate-Hosted Pb-Zn Deposits in the Central. *Remote Sens.* 12, 1239. doi: 10.3390/rs12081239.

Shearer, C. K., and Papike, J. J. (1999). Magmatic evolution of the Moon. *Am. Mineral.* 84, 1469–1494. doi: 10.2138/am-1999-1001.

Shearer, C. K., Sharp, Z. D., Burger, P. V., McCubbin, F. M., Provencio, P. P., Brearley, A. J., et al. (2014). Chlorine distribution and its isotopic composition in "rusty rock" 66095. Implications for volatile element enrichments of "rusty rock" and lunar soils, origin of "rusty" alteration, and volatile element behavior on the Moon. *Geochim. Cosmochim. Acta* 139, 411–433. doi: 10.1016/j.gca.2014.04.029.

Shi, C., and Wang, L. (2014). Incorporating spatial information in spectral unmixing: A review. *Remote Sens. Environ.* 149, 70–87. doi: 10.1016/j.rse.2014.03.034.

Shuai, T., Zhang, X., Zhang, L., and Wang, J. (2013). Mapping global lunar abundance of plagioclase, clinopyroxene and olivine with Interference Imaging Spectrometer hyperspectral data considering space weathering effect. *Icarus* 222, 401–410. doi: 10.1016/j.icarus.2012.11.027.

Smith, M. D., Bandfield, J. L., and Christensen, P. R. (2000). Separation of atmospheric and surface spectral features in Mars Global Surveyor Thermal Emission Spectrometer (TES) spectra. *J. Geophys. Res.* 105, 9589. doi: 10.1029/1999JE001105.

Soydan, H., Koz, A., and Düzgün, H. Ş. (2021). Secondary Iron Mineral Detection via Hyperspectral

# Supplementary Figures

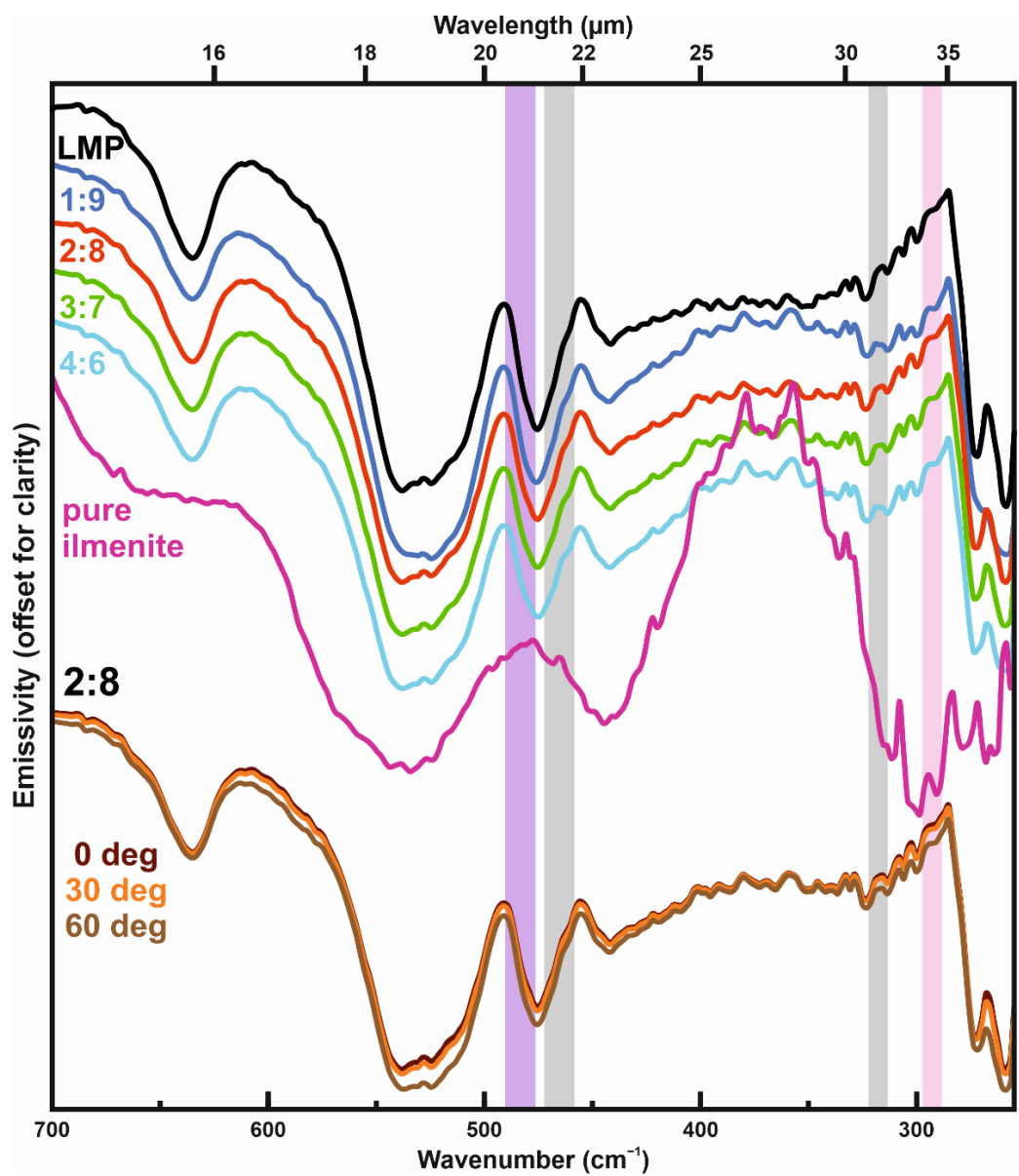

**Supplementary Figure 1.** The emissivity of ilmenite from the Arizona State University (ASU) spectral library (https://speclib.asu.edu/, ilmenite #463), equivalents of lunar mare basalts, and their various mixtures (1:9, 2:8, 3:7, and 4:6). The composition of the lunar mare basalt equivalents is the same as in the Section 5 text but normalized from 85% to 100%. The input spectra of olivine (fayalite), plagioclase (labradorite), and pyroxene (pigeonite) are also from the ASU spectral library. Various ilmenite and lunar mare basalt equivalent mixtures were simulated using a single-scattering albedo/Hapke model (Harris and Grindrod (2018). The mixtures are simulated for a solar incidence angle of 30°, but the differences between the 0–60° range are nearly negligible as demonstrated by the comparison of various angles for the 2:8 mixtures shown in the figure. The spectral ranges of four of the eight detectors are marked with rectangles: pale pink (288–297 cm$^{-1}$ = 33.69–34.71 µm: ilmenite), gray (313–322 cm$^{-1}$ = 31.03–31.97 µm and 458–472 cm$^{-1}$ = 21.18–21.82 µm: reference bands), and purple (476–490 cm$^{-1}$ = 20.39–21.01 µm: clinopyroxene). The ordinate axes do not have a numeric scale because the graphs are shifted for clarity.